\documentclass[journal]{IEEEtran}

\usepackage{amssymb}
\usepackage{graphicx}
\usepackage[font=scriptsize]{subcaption}
\usepackage[font=footnotesize]{caption}
\usepackage{ifpdf}

\usepackage{amsmath}
\usepackage{algorithmic}
\usepackage{array}
\usepackage{url}
\usepackage{cite}
\usepackage{fancyhdr}
\usepackage{mdwmath}
\usepackage{mdwtab}

\newtheorem{theorem}{Theorem}

\newtheorem{lemma}{Lemma}

\newtheorem{corollary}{Corollary}

\usepackage{epstopdf}
\usepackage[acronyms,nonumberlist,nopostdot,nomain,nogroupskip]{glossaries}

\newacronym{3gpp}{3GPP}{3rd Generation Partnership Project}
\newacronym{adc}{ADC}{Analog to Digital Converter}
\newacronym{dac}{DAC}{Digital to Analog Converter}
\newacronym{5g}{5G}{5th generation}
\newacronym{6g}{6G}{6th generation}
\newacronym{aimd}{AIMD}{Additive Increase Multiplicative Decrease}
\newacronym{am}{AM}{Acknowledged Mode}
\newacronym{amc}{AMC}{Adaptive Modulation and Coding}
\newacronym{aqm}{AQM}{Active Queue Management}
\newacronym{awgn}{AGWN}{Additive White Gaussian Noise}
\newacronym{balia}{BALIA}{Balanced Link Adaptation}
\newacronym{bdp}{BDP}{Bandwidth-Delay Product}
\newacronym{bf}{BF}{beamforming}
\newacronym{abf}{ABF}{analog beamforming}
\newacronym{dbf}{DBF}{digital beamforming}
\newacronym{hbf}{HBF}{hybrid beamforming}
\newacronym{cc}{CC}{Congestion Control}
\newacronym{cdf}{CDF}{Cumulative Distribution Function}
\newacronym{pdf}{pdf}{probability density function}
\newacronym{cn}{CN}{Core Network}
\newacronym{cqi}{CQI}{Channel Quality Information}
\newacronym{cp}{CP}{Control Plane}
\newacronym{csirs}{CSI-RS}{Channel State Information - Reference Signal}
\newacronym{dc}{DC}{Dual Connectivity}
\newacronym{rb}{RB}{Resource Block}
\newacronym{dce}{DCE}{Direct Code Execution}
\newacronym{dci}{DCI}{Downlink Control Information}
\newacronym{udp}{UDP}{User Datagram Protocol}
\newacronym{dl}{DL}{Downlink}
\newacronym{dmr}{DMR}{Deadline Miss Ratio}
\newacronym{dmrs}{DMRS}{DeModulation Reference Signal}
\newacronym{e2e}{E2E}{End-to-End}
\newacronym{ppp}{PPP}{Poisson Point Process}
\newacronym{si}{SI}{Study Item}
\newacronym{ecn}{ECN}{Explicit Congestion Notification}
\newacronym{edf}{EDF}{Earliest Deadline First}
\newacronym{enb}{eNB}{eNodeB}
\newacronym{epc}{EPC}{Evolved Packet Core}
\newacronym{es}{ES}{Edge Server}
\newacronym{cav}{CAV}{Connected and Autonomous Vehicle}
\newacronym{fdma}{FDMA}{Frequency Division Multiple Access}
\newacronym{fdd}{FDD}{Frequency Division Duplexing}
\newacronym{upa}{UPA}{Uniform Planar Array}
\newacronym{ula}{ULA}{Uniform Linear Array}
\newacronym{af}{AF}{array factor}
\newacronym[firstplural=Radio Access Technologies (RATs)]{rat}{RAT}{Radio Access Technology}
\newacronym[firstplural=Radio Access Technology (RTs)]{rt}{RT}{Radio Technology}
\newacronym{fs}{FS}{Fast Switching}
\newacronym{isd}{ISD}{inter-site distance}
\newacronym{ftp}{FTP}{File Transfer Protocol}
\newacronym{gnb}{gNB}{Next Generation Node Base}
\newacronym{harq}{HARQ}{Hybrid Automatic Repeat reQuest}
\newacronym{hetnet}{HetNet}{Heterogeneous Network}
\newacronym{hh}{HH}{Hard Handover}
\newacronym{hol}{HOL}{Head-of-Line}
\newacronym{ia}{IA}{Initial Access}
\newacronym{imt}{IMT}{International Mobile Telecommunication}
\newacronym{iot}{IoT}{Internet of Things}
\newacronym{los}{LOS}{line of sight}
\newacronym{lte}{LTE}{Long Term Evolution}
\newacronym{m2m}{M2M}{Machine to Machine}
\newacronym{mac}{MAC}{Medium Access Control}
\newacronym{mc}{MC}{Multi-Connectivity}
\newacronym{mcs}{MCS}{Modulation and Coding Scheme}
\newacronym{mec}{MEC}{Mobile Edge Cloud}
\newacronym{mi}{MI}{Mutual Information}
\newacronym{mimo}{MIMO}{Multiple Input Multiple Output}
\newacronym{mmwave}{mmWave}{millimeter wave}
\newacronym{mptcp}{MPTCP}{Multipath TCP}
\newacronym{mr}{MR}{Maximum Rate}
\newacronym{mss}{MSS}{Maximum Segment Size}
\newacronym{mtd}{MTD}{Machine-Type Device}
\newacronym{mtu}{MTU}{Maximum Transmission Unit}
\newacronym{nfv}{NFV}{Network Function Virtualization}
\newacronym{vnf}{VNF}{ Virtualization Network Function}
\newacronym{sdn}{SDN}{Software Defined Networking}
\newacronym{nlos}{NLOS}{non-line of sight}
\newacronym{nlosb}{NLOSb}{Building Non Line of Sight}
\newacronym{nlosv}{NLOSv}{Vehicle Non Line of Sight}
\newacronym{nr}{NR}{New Radio}
\newacronym{ofdm}{OFDM}{Orthogonal Frequency Division Multiplexing}
\newacronym{pdcch}{PDCCH}{Physical Downlonk Control Channel}
\newacronym{pdcp}{PDCP}{Packet Data Convergence Protocol}
\newacronym{pdsch}{PDSCH}{Physical Downlink Shared Channel}
\newacronym{pdu}{PDU}{Packet Data Unit}
\newacronym{pf}{PF}{Proportional Fair}
\newacronym{pgw}{PGW}{Packet Gateway}
\newacronym{phy}{PHY}{Physical}
\newacronym{pbch}{PBCH}{Physical Broadcast Channel}
\newacronym[plural=\gls{mme}s,firstplural=Mobility Management Entities (MMEs)]{mme}{MME}{Mobility Management Entity}
\newacronym{prb}{PRB}{Physical Resource Block}
\newacronym{pss}{PSS}{Primary Synchronization Signal}
\newacronym{pucch}{PUCCH}{Physical Uplink Control Channel}
\newacronym{pusch}{PUSCH}{Physical Uplink Shared Channel}
\newacronym{rach}{RACH}{Random Access Channel}
\newacronym{ran}{RAN}{Radio Access Network}
\newacronym{red}{RED}{Random Early Detection}
\newacronym{rf}{RF}{Radio Frequency}
\newacronym{rlc}{RLC}{Radio Link Control}
\newacronym{rlf}{RLF}{Radio Link Failure}
\newacronym{rrc}{RRC}{Radio Resource Control}
\newacronym{rrm}{RRM}{Radio Resource Management}
\newacronym{rr}{RR}{Round Robin}
\newacronym{rs}{RS}{Remote Server}
\newacronym{rsrp}{RSRP}{Reference Signal Received Power}
\newacronym{rss}{RSS}{Received Signal Strength}
\newacronym{rtt}{RTT}{Round Trip Time}
\newacronym{rw}{RW}{Receive Window}
\newacronym{rx}{RX}{Receiver}
\newacronym{sa}{SA}{standalone}
\newacronym{sack}{SACK}{Selective Acknowledgment}
\newacronym{sap}{SAP}{Service Access Point}
\newacronym{sch}{SCH}{Secondary Cell Handover}
\newacronym{scoot}{SCOOT}{Split Cycle Offset Optimization Technique}
\newacronym{sdma}{SDMA}{Spatial Division Multiple Access}
\newacronym{sinr}{SINR}{Signal to Interference plus Noise Ratio}
\newacronym{sm}{SM}{Saturation Mode}
\newacronym{snr}{SNR}{Signal to Noise Ratio}
\newacronym{son}{SON}{Self-Organizing Network}
\newacronym{ss}{SS}{Synchronization Signal}
\newacronym{srs}{SRS}{Sounding Reference Signal}
\newacronym{sss}{SSS}{Secondary Synchronization Signal}
\newacronym{tcp}{TCP}{Transmission Control Protocol}
\newacronym{tdd}{TDD}{Time Division Duplexing}
\newacronym{tdma}{TDMA}{Time Division Multiple Access}
\newacronym{tfl}{TfL}{Transport for London}
\newacronym{tm}{TM}{Transparent Mode}
\newacronym{prr}{PRR}{Packet Reception Ratio}
\newacronym{trp}{TRP}{Transmitter Receiver Pair}
\newacronym{tti}{TTI}{Transmission Time Interval}
\newacronym{ttt}{TTT}{Time-to-Trigger}
\newacronym{tx}{TX}{Transmitter}
\newacronym{ue}{UE}{User Equipment}
\newacronym{ul}{UL}{Uplink}
\newacronym{uml}{UML}{Unified Modeling Language}
\newacronym{um}{UM}{Unacknowledged Mode}
\newacronym{utc}{UTC}{Urban Traffic Control}
\newacronym{vm}{VM}{Virtual Machine}
\newacronym{rsrq}{RSRQ}{Reference Signal Received Quality}
\newacronym{rssi}{RSSI}{Received Signal Strength Indicator}
\newacronym{crs}{CRS}{Cell Reference Signal}
\newacronym{v2v}{V2V}{Vehicle-to-Vehicle}
\newacronym{v2i}{V2I}{Vehicle-to-Infrastructure}
\newacronym{v2n}{V2N}{Vehicle-to-Network}
\newacronym{v2x}{V2X}{Vehicle-to-Everything}
\newacronym{vn}{VN}{Vehicular Node}
\newacronym{dsrc}{DSRC}{Dedicated Short Range Communication}
\newacronym{ci}{CI}{context information}
\newacronym{voi}{VoI}{value of information}
\newacronym{gps}{GPS}{Global Positioning System}
\newacronym{qos}{QoS}{Quality of Service}
\newacronym{qoe}{QoE}{Quality of Experience}
\newacronym{ml}{ML}{Machine Learning}
\newacronym{ahp}{AHP}{Analytic Hierarchy Process}
\newacronym{lidar}{LIDAR}{Light Detection and Ranging}
\newacronym{sumo}{SUMO}{Simulation of Urban MObility}
\newacronym{wave}{WAVE}{Wireless Access in Vehicular Environment}
\newacronym{c-its}{C-ITS}{Connected Intelligent Transportation System}
\newacronym{dash}{DASH}{Dynamic Adaptive Streaming over HTTP}
\newacronym{http}{HTTP}{HyperText Transfer Protocol}
\newacronym{ntn}{NTN}{non-terrestrial network}
\newacronym{ntc}{NTC}{non-terrestrial communication}
\newacronym{haps}{HAPS}{High Altitude Platform Station}
\newacronym{hap}{HAP}{High Altitude Platform}
\newacronym{leo}{LEO}{Low Earth Orbit}
\newacronym{meo}{MEO}{Medium Earth Orbit}
\newacronym{geo}{GEO}{Geostationary Earth Orbit}
\newacronym{uav}{UAV}{Unmanned Aerial Vehicle}
\newacronym{nsat}{nSAT}{Nanosatellite}
\newacronym{ehf}{EHF}{extremely high-frequency}
\newacronym{ioe}{IoE}{Internet of Everyone}
\newacronym{gan}{GaN}{Gallium Nitride}
\newacronym{aoi}{AoI}{Area of Interest}
\newacronym{vb}{VB}{vertical beam}
\newacronym{tb}{TB}{tilted beam}
\newacronym{gu}{GU}{ground user}
\newacronym{hpbw}{HPBW}{half-power beamwidth}
\newacronym{pae}{PAE}{power-added efficiency}
\newacronym{2d}{2D}{two-dimensional}
\usepackage{hyperref}
\usepackage{mathtools}
\DeclarePairedDelimiter\ceil{\lceil}{\rceil}

\usepackage{tikz}
\usepackage{pgfplots}
\pgfplotsset{compat=newest} 
\pgfplotsset{plot coordinates/math parser=false} 
\newlength\fheight
\newlength\fwidth
\usetikzlibrary{plotmarks,patterns,decorations.pathreplacing,backgrounds,calc,arrows,arrows.meta,spy,matrix}
\usepgfplotslibrary{patchplots,groupplots}
\usepackage{tikzscale}
\usepackage{hyperref}
\usepackage{siunitx}

\usepackage{multirow}
\usepackage{tkz-kiviat}
\usepackage{tikz-qtree}
\usetikzlibrary{trees} 
\usepackage{siunitx}
\usetikzlibrary{fadings}

\tikzfading[name=middle,
            top color=transparent!100,
            bottom color=transparent!100,
            middle color=transparent!20]

\usetikzlibrary{arrows,automata,calc,shapes, positioning,shadows,shadows.blur,shapes.geometric}

\begin{document}

\title{On\hspace{0.19cm}the\hspace{0.19cm}Beamforming\hspace{0.19cm}Design\hspace{0.19cm}of\hspace{0.19cm}Millimeter\hspace{0.19cm}Wave\hspace{0.19cm}UAV\hspace{0.19cm}Networks:\hspace{0.19cm}Power\hspace{0.19cm}vs.\hspace{0.19cm}Capacity\hspace{0.19cm}Trade-Offs}

\author{{{Yang Wang},~\IEEEmembership{Student Member, IEEE},
        {Marco Giordani},~\IEEEmembership{Member, IEEE},\\
        {Michele Zorzi},~\IEEEmembership{Fellow, IEEE}}
        \thanks{Yang Wang is with the Beijing University of Posts and Telecommunications, Beijing, China (email: ddf@bupt.edu.cn). \newline
        \indent Marco Giordani and Michele Zorzi are with the Department of Information Engineering, University of Padova, Padova, Italy (email: \{giordani,zorzi\}@dei.unipd.it).}
}



\maketitle

\begin{abstract}
The millimeter wave (mmWave) technology enables unmanned aerial vehicles (UAVs) to offer broadband high-speed wireless connectivity in fifth generation (5G) and beyond (6G) networks. 
However, the limited footprint of a single  UAV implementing analog beamforming (ABF) requires multiple aerial stations to operate in swarms to provide  ubiquitous network coverage, thereby posing serious constraints in terms of battery power consumption and swarm management. A possible remedy is to investigate the concept of hybrid beamforming (HBF) transceivers, which use a combination of analog beamformers as a solution to achieve higher flexibility in the beamforming design. 
This approach permits multiple ground users to be served simultaneously by the same UAV station, despite involving higher energy consumption in the radio frequency (RF)
domain than its ABF counterpart.
This paper presents a tractable stochastic analysis to characterize the downlink ergodic capacity and power consumption of UAV mmWave networks in an urban scenario, focusing on the trade-off between ABF and HBF  architectures. A multi-beam coverage model is derived as a function of several UAV-specific parameters, including the number of UAVs, the deployment altitude, the antenna configuration, and the beamforming design. 
Our results, validated by simulation, show that, while ABF achieves better ergodic capacity at high altitudes, an HBF configuration with multiple beams, despite the use of more power-hungry RF blocks, consumes less power all the time with limited capacity degradation.
\end{abstract}

\begin{picture}(0,0)(-33,-430)
\put(0,0){
\put(0,0){This paper has been submitted to IEEE for publication. Copyright may be transferred without notice.}}
\end{picture}


\IEEEpeerreviewmaketitle

\section{Introucion}


\IEEEPARstart{W}{hile} \gls{lte} networks are already successfully penetrating new markets, and with \gls{5g} deployments ready for global commercial roll-out, the  more stringent traffic requirements of future wireless applications (which will reach 1 zettabyte/month in 2028~\cite{khan2016multi}) are driving the research community to discuss new services  and enabling technologies towards \gls{6g} systems~\cite{giordani2020towards}. 
Notably, not only will current terrestrial infrastructures facing greater capacity demands  not guarantee the required \gls{qos}, but they will also show vulnerability in emergency situations (such as disaster relief/recovery), a research challenge that communication standards have, so far, relegated to the very bottom, if not entirely ignored~\cite{chaoub20206g}.




To address this issue, 6G research is  focusing on the development of \glspl{ntn} where air/spaceborne stations like \glspl{uav}, \glspl{hap}, and satellites  assist terrestrial infrastructures in promoting flexible and cost-effective global connectivity~\cite{giordani2019non,giordani2020satellite}.
Thanks to their high mobility, versatility, and low cost, \glspl{uav}, in particular,  play a key role in providing network service recovery in devastated region, enhancing public safety, and handling emergency situations\cite{li2018uav,7470933,8764406,8758340}.

In combination with the \gls{mmwave} technology, whose huge available bandwidth can offer multi-Gbps data rates~\cite{7959169},  UAVs may also serve as aerial relays and base stations to support wireless services in high-traffic scenarios, e.g.,  to assist backhaul operations when terrestrial towers are overloaded~\cite{8482308}, and/or to relay large amounts of sensor data in hot-spot areas from multiple \glspl{gu}~\cite{8999435}. 
However, mmWave communications may incur severe path loss and sensitivity to blockage, especially considering the very long transmission distances involved~\cite{giordani2019tutorial}, which may result in \gls{qos} degradation. 
In these regards,  massive \gls{mimo} techniques have been introduced to improve reliability and spectral efficiency through beamforming~\cite{7959169}. 
The short wavelength of \gls{mmwave} signals, indeed, permits multiple antenna elements to be placed into a small \gls{uav} to form  a concentrated beam pattern towards a specific direction, thus providing array gains and reduced co-channel interference~\cite{sun2014mimo}.

In the UAV context, the limited energy and payload resources of aerial systems make it difficult, if not impossible, to realize full-blown \gls{dbf} which, while  potentially enabling the transceiver to direct beams at infinitely many directions, requires a dedicated \gls{rf} chain for each antenna element~\cite{7961162}. 
An \gls{abf} structure, in turn, adopts the simplest electronic components and requires a single \gls{rf} chain, thus theoretically representing the most desirable option to achieve low power consumption~\cite{xiao2020unmanned}. 
The limited flexibility of \gls{abf}, however, means that UAVs can only beamform in one direction at a time, and forces multiple aerial stations to be deployed in formation to provide ubiquitous network coverage to GUs: this may incur significant energy consumption for propulsion and hovering as the swarm grows in size, thereby posing severe power management constraints.
In this context, the research community is leaning towards the development of \gls{hbf}, which realizes a simple transceiver design by combining \gls{rf} analog beamformers and low-dimensional baseband digital beamformers~\cite{sohrabi2017hybrid,molisch2017hybrid}. In particular, HBF enables simultaneous transmission of multiple data streams from the same UAV station, and makes it possible to reduce the UAV swarm size and its relative cost compared to ABF.
Despite these premises, however, few contributions have been devoted to the evaluation of the performance of \gls{hbf} in a UAV scenario, which in turn represents a  timely research~issue.

Following this rationale, in this paper we provide the first analytical model to jointly evaluate and compare the ergodic capacity and power consumption of \gls{abf} and \gls{hbf} architectures for \gls{uav} networks. 
The main novelty of our work can be summarized as follows:
\begin{itemize}
	\item We provide a mathematical and tractable expression  for the ergodic capacity of UAVs operating at \gls{mmwave} frequencies based on stochastic geometry. Compared to previous analyses, which employ a specific beamforming design (e.g.,~\cite{ravi2016downlink,chetlur2017downlink}), in this work we compare the performance of  the \gls{abf} and \gls{hbf} configurations as a function of  the UAV deployment altitude, the antenna architecture, and the density of GUs. 
	For these two types of beamforming strategies, two approximate antenna patterns (based on a flat-top model that accounts for both direct and tilted beams) are proposed to achieve a good balance between accuracy and analytical tractability. 

	\item We provide an analytical expression for the power consumed by UAVs when establishing \gls{abf} and \gls{hbf} transmissions. Our model accounts for both the   hovering power, which  is required to maintain the UAV aloft and enable its mobility, as well as the power consumed by each MIMO electronic component for communication. We discuss the impact of the number of \gls{rf} chains and antenna elements on the beamforming design, the UAV deployment altitude, and the resolution of the \gls{dac}, i.e., the most power-hungry hardware block in the transmitter. 

	\item Based on the above-mentioned relations, while existing analyses tend to separate coverage- (e.g.,~\cite{yi2019modeling,boschiero2020coverage}) and power-related (e.g.,~\cite{di2015energy,naqvi2018energy}) performance, we provide a new paradigm to jointly evaluate the coverage vs. power consumption trade-off for mmWave-assisted UAV networks. Our theoretical model is validated via realistic Monte Carlo simulations, so as to include many more details than would be possible via analytical evaluations
while accounting for realistic channel implementations.
Despite the common belief  that HBF suffers from higher energy consumption than  ABF, we demonstrate that HBF with two or three simultaneous beams always outperforms an ABF strategy at low altitude in terms of both power efficiency and capacity, especially when increasing the antenna array size. Conversely, ABF exhibits better capacity at higher altitudes, even though HBF's degradation is limited. Specifically, we demonstrate that the user's density qualifies as the most crucial factor in the capacity performance.

\item We provide guidelines on the optimal beamforming strategy to adopt to minimize power consumption  while not sacrificing  spectral efficiency. 
We show that, with the increasing UAV altitude, the system capacity has a peak value that depends on the density of \glspl{gu} and the antenna array size, above which the system would consume more power.

\end{itemize}

The remainder of this paper is organized as follows.  Related works on  UAV \gls{mmwave} networks  are discussed in Sec.~\ref{sec:related}. Sec.~\ref{sec:system_model} describes our system model, including the scenario, channel, antenna, and power consumption models for both ABF and HBF architectures. In Sec.~\ref{sec:capacity} and Sec.~\ref{sec:power} we derive the expressions of the power consumption and  ergodic capacity for ABF and HBF, respectively. Sec.~\ref{sec:evaluation} reports our system parameters, 
 validates our theoretical framework through simulations, and presents our main findings and results. Finally, conclusions and suggestions for future work are provided in Sec.~\ref{sec:conclusions}.

\section{Related Work}
\label{sec:related}

UAV-assisted mmWave networks have been extensively studied in the literature. 
Xiao \emph{et al.}, for example, reviewed the main opportunities and challenges associated with high-frequency \gls{uav} operations~\cite{xiao2016enabling},
Xia \emph{et al.} discussed applications of mmWaves in \gls{uav}-enabled public safety scenarios~\cite{xia2019millimeter}, while Zhao \emph{et al.} evaluated the performance of integrated flight control and channel tracking for mmWave aerial links~\cite{zhao2018channel}.
More recent works are now focusing on analyzing the coverage and data rate performance, beamforming design, and power consumption of the network, as discussed in the following~paragraphs.
\smallskip

\paragraph{Coverage/rate performance}
Stochastic geometry has emerged as a tractable approach to study the coverage performance of wireless systems. 
In the UAV context, papers~\cite{chetlur2017downlink,azari2017coverage,ravi2016downlink,liu2018performance} derived lower and upper bounds for the coverage probability and the area spectral efficiency of UAV networks, while in our related work~\cite{boschiero2020coverage} we identified the trade-offs to be considered to maximize the coverage performance as a function of the drone density, height, and antenna patterns.
More recently, the article in \cite{8856258} investigated the coverage performance of UAV  mmWave networks when UAVs are assumed to be distributed according to a homogeneous \gls{ppp} and users are modeled as a  Poisson Cluster Process (PCP, e.g., Thomas cluster processes or Matern cluster processes). 
For the proposed system, simulation results indicated that there exists an optimal altitude that maximizes the coverage probability, and that the effect of the thermal noise and non-line of sight transmissions can be omitted. 
Related work~\cite{8876702} adopted a similar model for the deployment of UAVs and \glspl{gu}, and formulated a mathematical expression for the average uplink throughput to identify the optimal time division multiplexing between the downlink and uplink phases. 
Similarly, the authors in  \cite{colpaert2018aerial} compared the performance of LTE and mmWave  networks for serving flying UAVs. The results demonstrated that a significant improvement of the UAV coverage probability is possible using mmWaves, even though inter-cell interference remains a key performance challenge at high altitudes.
Another approach was considered in~\cite{9115248}, where the authors provided a closed-form expression for the maximum achievable rate for a full-duplex UAV scenario, and proposed an alternating interference suppression algorithm to jointly design the beamforming vectors and the power control variables. Moreover, Xiao \emph{et al.}~\cite{8907440} formulated a non-convex problem to maximize the achievable data rate, subject to user requirements, UAV positions, and beamforming vector~constraints.
\smallskip

\paragraph{Energy consumption}
Despite demonstrating promising coverage performance, there are still several challenges associated to UAV-enabled networks, ranging from energy limitations to optimal 3D deployment. In the literature, Mozaffari \emph{et al.}~\cite{mozaffari2015drone} determined the optimal UAV altitude to minimize the transmitted power, while the authors in~\cite{di2015energy} proposed a coverage-aware path planning method to reduce energy consumption while satisfying data rate and resolution constraints. 
Several other related works, e.g.,~\cite{yang2018energy,zeng2019energy}, have also analyzed the energy trade-offs associated with UAV propulsion, which involves significantly higher power consumption compared to conventional (static) terrestrial stations.

In this perspective, the utilization of the \gls{mmwave} technology for drone-assisted communications also resulted in higher power utilization for electronic components like \glspl{dac}, phase shifters and combiners, and power amplifiers, which are required to process a large number of antenna outputs and very wide bandwidths.
In a recent work~\cite{orhan2015low}, the authors explored how DAC resolution and bandwidth affect the total power consumption, while~\cite{buzzi2018energy} studied the energy performance of  mmWave systems as a function of different pre-coding and post-coding beamforming structures.
However, most literature is referred to a cellular-type scenario, while only a few papers have investigated energy efficiency for UAV networks~\cite{zeng2019energy,naqvi2018energy}, which therefore represents a timely topic for further research.
\smallskip

\paragraph{Beamforming design}
Sharp beamforming is necessary to support UAV communications at \glspl{mmwave}. 
Along these lines, Miao \emph{et al.}~\cite{miao2019position} first formalized a robust beamforming optimization problem to maximize network coverage under the constraint of strong interference. Similarly, the authors in~\cite{yang2019beam} explored how to design the optimal beam according to the drone flying range, while paper~\cite{miao2019lightweight} presented an innovative position-based lightweight beamforming technology to enhance broadcasting   communications through UAVs.
More recently, Li \emph{et al.}~\cite{li2020reconfigurable} proposed to jointly optimize UAV trajectory and beamforming design through reconfigurable intelligent surfaces, thus improving  propagation environment and enhancing communication quality.
While full \gls{dbf} cannot be adopted on UAVs because of the high cost and energy consumption, \gls{hbf} solutions have been proposed to improve communication performance while minimizing power utilization, e.g.,~\cite{yu2019low}.  A similar approach was followed in~\cite{ren2019machine}, even though the literature does not carefully demonstrate whether (and in which form) \gls{hbf} outperforms a pure \gls{abf} configuration from  both a coverage and an energy efficiency point of view. 
\smallskip

Our work differs from the prior art in that we compare different beamforming architectures, optimizing both power consumption and coverage performance together. Moreover, our power model does not only consider hovering operations, but it also characterizes the energy impact for communication in the UAV~context.

\section{System Model}
\label{sec:system_model}

In this section we present our system scenario (Sec.~\ref{ssec:scenario}), channel (Sec.~\ref{ssec:channel}), antenna (Sec.~\ref{ssec:antenna}), and power (Sec.~\ref{ssec:power}) models that are considered in our study. 

\begin{figure}[t!]
	\centering
    \setlength{\belowcaptionskip}{-0.43cm}
	\includegraphics[width=0.99\columnwidth]{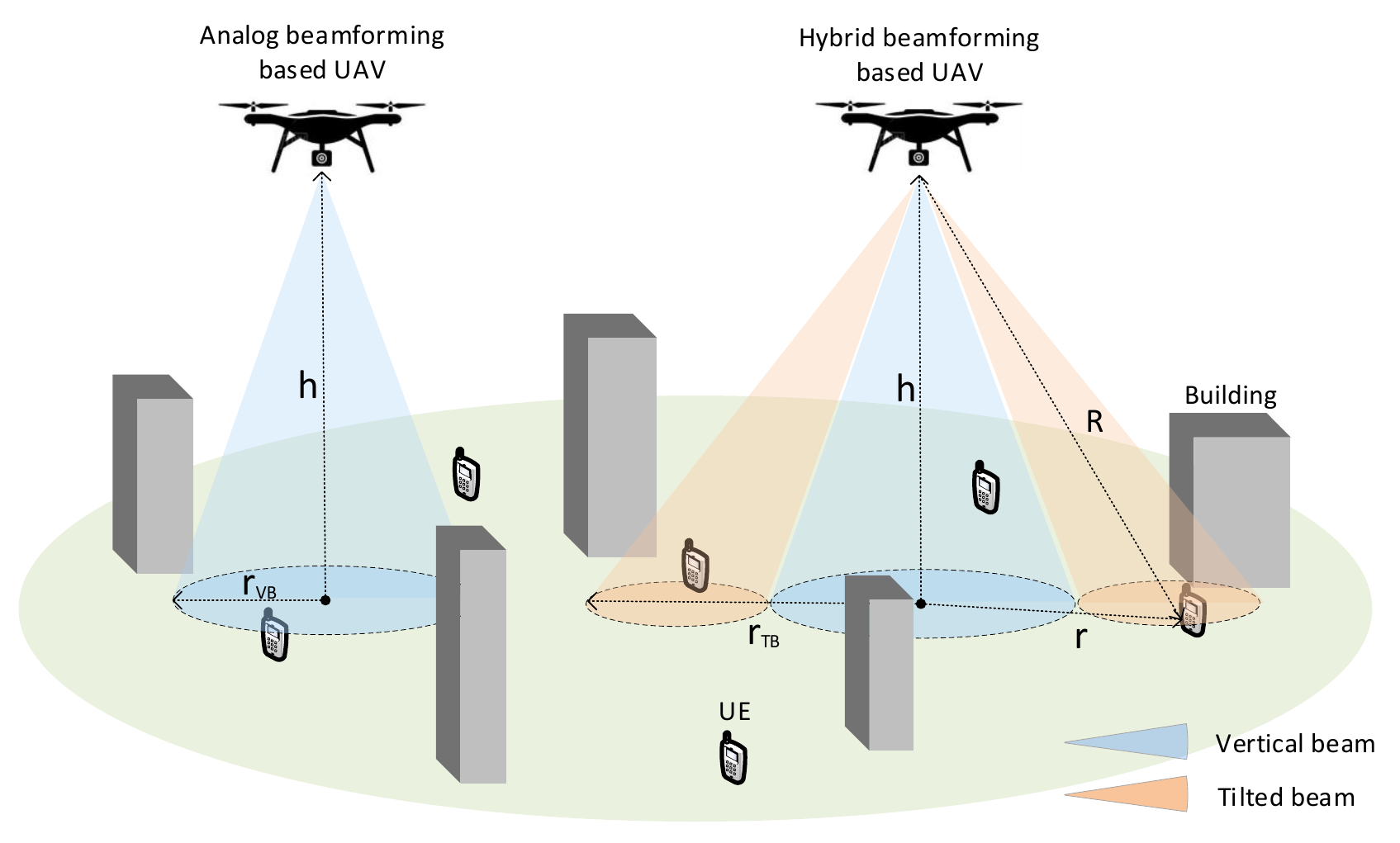}
	\caption{UAV-assisted system model. UAVs can form vertical and/or tilted beams, depending on the beamforming architecture (ABF or HBF, respectively) they adopt.}
	\label{fig:scenario}
\end{figure}

\subsection{Scenario Description}
\label{ssec:scenario}

As shown in Fig.~\ref{fig:scenario}, we consider a UAV-assisted urban scenario in which mmWave-enabled swarms of drones acting as aerial base stations are deployed at altitude $h$ to provide communication services on the ground. Furthermore, we assume that \glspl{gu} are placed across a circular \gls{aoi} to form a \gls{2d} homogeneous \gls{ppp} $\Phi_u$ with density $\lambda _{u}$.
We assume that both UAVs and GUs communicate through directional beams whose width, shape, and structure depend on the beamforming strategy they adopt, as we will specify in Sec.~\ref{ssec:antenna}.
We then call $r$ the \gls{2d} distance between the projection
 of the UAV on the user plane and a generic GU in the \gls{aoi}, and $R=\sqrt{r^2+h^2}$ the distance between the
reference UAV and the \gls{gu}.

\subsection{Channel Model}
\label{ssec:channel}
\subsubsection{Line-of-sight probability}
In the urban scenario, air-to-ground communication links may be blocked by environmental structures such as buildings or vegetation. Therefore, it is essential to distinguish between \gls{los} and \gls{nlos} propagation (denoted respectively with subscripts $L$ and $N$ throughout the paper). 
The International Telecommunication Union (ITU) presents a precise expression for the \gls{los} probability~\cite{al2014}, which is estimated by ray-tracing techniques using data from building and terrain databases. For the purpose of mathematical tractability, in this paper we adopt a simplified equation, proposed in prior work~\cite{al2014optimal} and universally adopted in the literature, in which the \gls{los} probability $p_{L}(r)$ is modeled, as a function of the GU-UAV distance $R=\sqrt{r^2+h^2}$, as a modified Sigmoid function (S-curve), i.e.,
\begin{align}
p_{L}(r)=\frac{1}{1+p\: \mathrm{exp}\left[-q\left(\arctan\Big(\frac{h}{\sqrt{r^2+h^2}}\Big)\frac{180}{\pi}-p\right)\right]},
\label{eq:p_l}
\end{align}
where $\arctan({h}/\sqrt{r^2+h^2})$ is the elevation angle of the UAV, and $p$ and $q$ are the S-curve parameters which depend on the environment (suburban, urban, dense urban, high-rise urban). Consequently, the \gls{nlos} probability is given by $p_{N}(r)=1-p_{L}(r)$.
By the thinning theorem of \glspl{ppp}, we can thus distinguish two independent PPPs for the \gls{los} and \gls{nlos} GUs, i.e., $\Phi_{u,L}\subseteq\Phi_{u}$ and  $\Phi_{u,N}\subseteq\Phi_{u}$, respectively, of intensity measures $\lambda_{u, L}=p_L(r)\lambda_u$ and $\lambda_{u, N}=p_N(r)\lambda_u$.

\subsubsection{Path gain}
We consider the widely adopted power-law model for the path gain~\cite{mozaffari2016unmanned}, i.e.,
\begin{align}
\ell_{i}(r)=\eta_{i}R^{-\kappa_{i}}=\eta_{i}(\sqrt{r^2+h^2})^{-\kappa_{i}}, \quad i\in \{L,\,N\}.
\label{eq:ell}
\end{align}
In Eq.~\eqref{eq:ell},  $\kappa_{i}$ is the path loss exponent and $\eta_{i}$ is the path loss gain at unit distance. Due to the signal propagating in \gls{nlos} suffering from shadowing and reflection effects from obstacles, it holds $\eta_{\mathrm{L}} > \eta_{\mathrm{N}}$ and $\kappa_{\mathrm{L}} < \kappa_{\mathrm{N}}$.

\subsubsection{Small-scale fading}
Even though measurements conducted in the mmWave environment show a relatively small impact of fading on propagation~\cite{6515173}, the effect of small-scale fading due to reflectivity and scattering from common objects may not be negligible in the UAV scenario, especially when directional beamforming is applied~\cite{zhu2018secrecy}. 
In this paper, small-scale fading is modeled as a Nakagami-m random variable $\gamma_i$ of parameters $m_i$ and $\Omega_i$, $i\in\{L,N\}$, which is typically adopted to represent both \gls{los} and \gls{nlos} air-to-ground channels, whose \gls{cdf} is given by~\cite{nakagami1960m}:
\medmuskip=2mu
\thickmuskip=2mu
\begin{align}
&F(x;m_i,\Omega_i)=\mathbb{P}[\gamma_i(m_i,\Omega_i)\leq x]= \notag \\
&P\left(m_i,\frac{m_i}{\Omega_i}x^2\right)=\frac{2m_{i}^{m_{i}}x^{2m_{i}-1}}{\Omega_i^{m_{i}} \Gamma (m_{i})}\ \mathrm{exp}\left(\frac{-m_{i}x^{2}}{\Omega _{i}}\right).
\label{eq:nakagami}
\end{align}
In Eq.~\eqref{eq:nakagami},  $\Omega _{i}$ is the spread factor, $m_{i}$ is the shape factor, and $\Gamma (\cdot)$ is the Gamma~function. 
\medmuskip=6mu
\thickmuskip=6mu

\subsection{Antenna Model}
\label{ssec:antenna}
In this paper we assume that UAVs are equipped with \glspl{upa} of $N_T=N_x\times N_y$ antenna elements, separated in the horizontal and vertical dimension by $d_{x}$ and $d_{y}$, respectively, providing gain by beamforming. For simplicity, we assume square \glspl{upa}, i.e., $N_x=N_y$,  and $d_{x}=d_{y}=d$. Further, we consider a flat-top model to characterize the pattern of the directional beamforming. 
UAVs implementing \gls{abf} shape the beam through a single \gls{rf} chain for all the antenna elements, and can form a single \gls{vb} perpendicular to the ground (the blue beams in Fig.~\ref{fig:scenario}). In this case, the processing is performed in the analog domain, thus transmitting/receiving in only one direction at a time~\cite{giordani2019tutorial}.
UAVs implementing \gls{hbf}, in turn, use $N_{RF}$ RF chains (with $N_{RF} \leq N_T$) that allow to produce, besides the analog \gls{vb}, $N_{D}-1$ additional parallel \glspl{tb}, with $N_D\leq N_{RF}$, towards the ground (the orange beams in Fig.~\ref{fig:scenario}). This permits the transceiver to transmit/receive in $N_D$ directions simultaneously, thus improving flexibility compared to ABF, at the cost of a higher power consumption.

\begin{lemma}\label{lemma: antenna gain}
Based on the above assumptions, the main lobe directivity gain $G$ for a UPA produced by beamforming can be expressed~as
\begin{equation}
G=N_{x}\cdot N_{y}.
\label{eq:bf_gain}
\end{equation}

\textit{Proof:} For \glspl{upa}, the \glspl{af} of the arrays in the x- and y-directions, for any pair of vertical and horizontal angles $(\theta ,\phi)$, $\forall \theta \in [0,\pi],\: \forall \phi \in [-\pi,\pi]$, can be written as 
\begin{align}
\mathrm{AF}(\theta ,\phi)=& \sum_{n_{y}=1}^{N_{y}} \left[ \sum_{n_{x}=1}^{N_{x}} e^{j(2\pi /\lambda)[(n_{x}-1)d_{x}\sin\theta\cos\phi]}\right] \notag \\
& \cdot  e^{j(2\pi /\lambda)[(n_{y}-1)d_{y}\sin\theta\sin\phi]},
\end{align}
where $\lambda$ is the wavelength.
Inspired by \cite{dabiri2020analytical} and \cite{yu2017coverage}, which derived the instantaneous directivity gain of \glspl{ula} transmitters, the array gain of \glspl{upa} can be recast as
\begin{align}
G_{t}(\Delta\theta,\Delta\phi)&= N_{x} N_{y}\left[\mathrm{AF}(\Delta\theta ,\Delta\phi)\right]^2 \notag \\
&\overset{(a)}{=}\left[ \frac{\sin ^{2}(\pi N_{x}\psi _{x})}{N_{x}\sin ^{2}(\pi \psi _{x})} \right]  \left[  \frac{\sin ^{2}(\pi N_{y}\psi _{y})}{N_{y}\sin ^{2}(\pi \psi _{y})} \right],   
\end{align}
where
\begin{align}
&\psi _{x}=\frac{d_{x}}{\lambda}\sin(\Delta\theta)\cos(\Delta\phi ), \notag \\
&  \psi _{y}=\frac{d_{y}}{\lambda}\sin(\Delta\theta)\cos(\Delta\phi ), 
\end{align}
and  $(a)$ follows from \cite{balanis2016antenna}. 
Assuming that the optimal beamforming vector is used, i.e., $\Delta\theta=\Delta\phi=0$, the antenna gain is then given~by
\begin{align}
G=G_{t}(0,0)=N_{x}\cdot N_{y}.  
\end{align}
\hfill $\blacksquare$
\end{lemma}

\subsection{Power Model}
\label{ssec:power}
In our model, we consider  the power for hovering, which maintains the UAV aloft and permits its mobility, and for~communicating to the GUs in both ABF and HBF~configurations.

\subsubsection{Hovering Power}
The power consumption model of multi-rotor UAVs include three components \cite{7991310}: (i) induced power ($P_{\mathcal{I}}$), which produces thrust by propelling air downwards, (ii) profile power ($P_{\mathcal{P}}$), which overcomes the rotational drag of rotating propellers, and (iii) parasite power ($P_{par}$), which resists body drag between the UAV and the wind. These components are expressed as
\medmuskip=4mu
\thickmuskip=4mu
\begin{align}
&P_{\mathcal{I}}(T,V_{vert})=k_{1}T\left[\tfrac{V_{vert}}{2}+\sqrt{(\tfrac{V_{vert}}{2})^{2}+\tfrac{T}{k_{2}^{2}}}\right],\\
&P_{\mathcal{P}}(T,V_{air})=c_{2}\sqrt{T^3}+c_{3}(V_{air}\cos\alpha_a)^{2}\sqrt{T},\\
&P_{par}(V_{air})=c_{4}V_{air}^{3},\\
&T=\sqrt{[mg-(c_{5}(V_{air}\cos\alpha)^{2}+c_{6}T)]^{2}+(c_{4}V_{air}^{2})^{2}},
\end{align}
\medmuskip=6mu
\thickmuskip=6mu
where $T$, $mg$, and $V_{vert}$ are the UAV thrust, gravity, and vertical speed, $V_{air}$  is the horizontal airspeed, and $\alpha_a$ is the angle of attack. $k_{1}$, $k_{2}$, $c_{2}$, $c_{3}$, $c_{4}$ and $c_{5}$ are related constant parameters. When the vertical speed and the horizontal airspeed are equal to zero (e.g., when considering UAVs hovering at fixed locations, as assumed in this study), the hovering power consumption can be expressed as
\begin{align}
P_{hover}&=P_{\mathcal{I}}(T,0)+P_{\mathcal{P}}(T,0)+P_{par}(0)\notag \\
&=(c_{1}+c_{2})(mg)^{3/2},
\label{eq:hover}
\end{align}
where $c_{1}=k_{1}/k_{2}$. The hovering power consumption determines the maximum UAV endurance, range, and flight speed, which, based on Eq.~\eqref{eq:hover}, mainly depends on the UAV gravity.

\begin{figure}[t!] 
  \begin{minipage}[t]{.45\textwidth}
    \centering
	\includegraphics[width=3in]{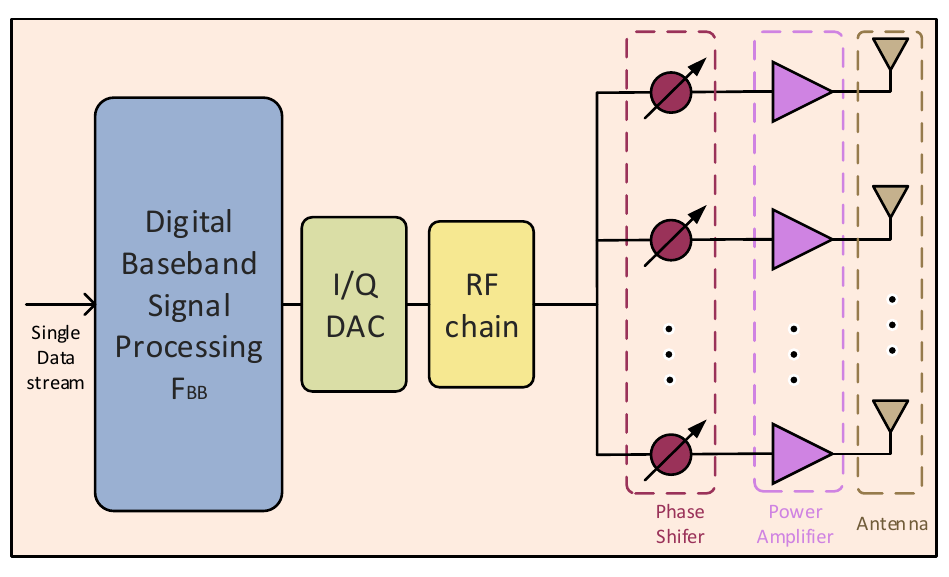}
	\caption{ABF architecture with single RF chain.}
    \vspace{0.53cm}
	\label{abf}
  \end{minipage}\\
  \begin{minipage}[t]{.45\textwidth} 
    \centering
    \centering
	\includegraphics[width=3in]{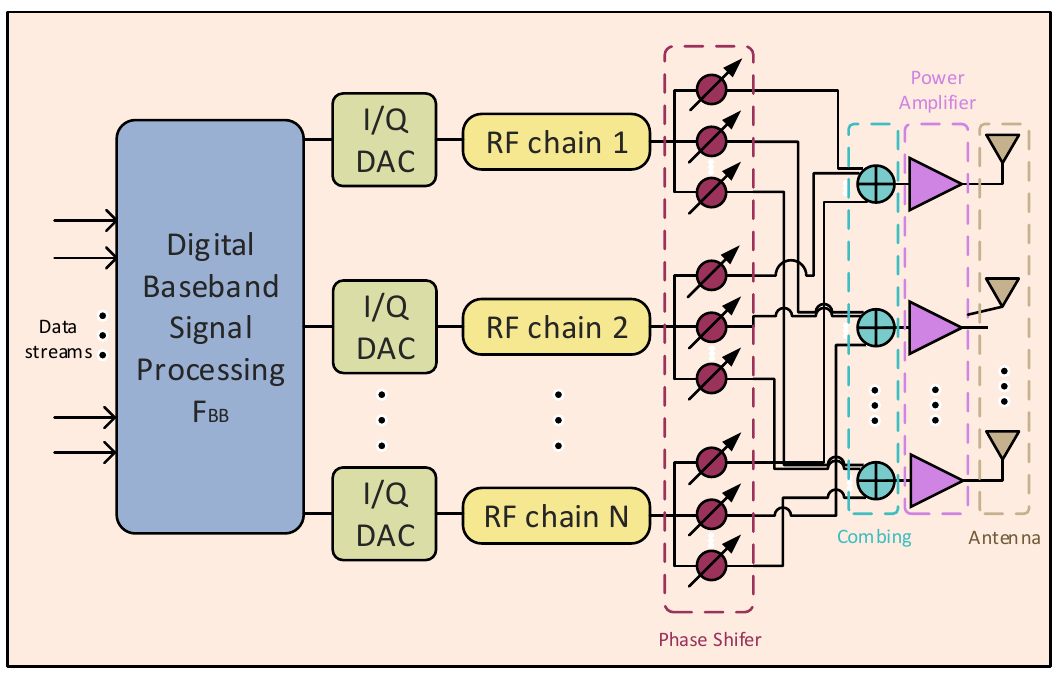}
	\caption{HBF architecture with multiple RF chains.}
	\label{hbf}
  \end{minipage}
\end{figure}%



\subsubsection{Communication Power}

The ABF and HBF architectures are shown in Figs. \ref{abf} and \ref{hbf}, respectively. The total static power consumption of each scheme required for communication towards the ground can be evaluated by the following expressions~\cite{8057288}:
\medmuskip=3mu
\thickmuskip=3mu
\begin{align}
&P_{\rm ABF}=P_{PA}+2P_{DAC}+P_{RF}+P_{SP}+N_{T}(P_{PS}),\label{eq:p_abp}\\[5pt]
&P_{\rm HBF}=P_{PA}+N_{RF} \left(2P_{DAC}+P_{RF} +P_{SP} \right)\notag\\
& \qquad \quad \:  +N_{T}(N_{RF}P_{PS}+P_{C}),
\label{eq:p_hbp}
\end{align}
\medmuskip=6mu
\thickmuskip=6mu

where $N_T$ is the total number of antenna elements in the UAV's array, $N_{RF}$ is the number of RF chains, and $P_{PA}$, $P_{PS}$, $P_{SP}$, $P_{C}$ represent the power consumptions of the  power amplifier, phase shifter, splitter, and combiner, respectively.
In Eqs.~\eqref{eq:p_abp} and \eqref{eq:p_hbp},  $P_{RF}$ and $P_{DAC}$ identify the power consumption of the RF chain and DAC, respectively. 
We consider an RF phase shifting approach, which modifies the phase on the RF path directly. Moreover, we deploy DACs with a binary-weighted current-steering topology which does not involve buffering, thus allowing high-speed conversions~\cite{8333733}. Hence, power consumption can be evaluated respectively as 
\medmuskip=3mu
\thickmuskip=3mu
\begin{align}
&P_{RF}=P_{M}+P_{LO}+P_{LPF}+P_{H}+P_{BBamp},\\
&P_{DAC}=1.5\times 10^{-5}\cdot 2^{b_{DAC}}+9\times 10^{-12}\cdot b_{DAC}\cdot F_{S},
\end{align}
\medmuskip=6mu
\thickmuskip=6mu
where $P_{M}$, $P_{LO}$, $P_{LPF}$, $P_{H}$, $P_{BBamp}$ are the power consumption of mixer, local oscillator, low pass filer, $90$-degree hybrid coupler with buffer, and base-band amplifier, while $F_{S}$ and $b_{DAC}$ are the DAC sampling frequency and resolution, respectively.
	
Overall, considering both UAV hovering and communication modules, the total power consumption for a UAV-assisted \gls{mmwave} transmitter can be expressed by
\begin{align}
&P_{\rm tot}^{\rm ABF}=P_{ABF}+P_{hover}, \label{eq:p-tot-abf}\\
&P_{\rm tot}^{\rm HBF}=P_{HBF}+P_{hover}.\label{eq:p-tot-hbf}
\end{align}

\section{Power Consumption Analysis}
\label{sec:power}
In this section, we first calculate the main lobe \gls{hpbw} of the VB and TBs (Sec.~\ref{ssec:coverage_model}), then we derive an analytical expression for the power consumed by UAVs implementing ABF or HBF  to provide connectivity to a certain \gls{aoi} of size $S_{\rm tot}$ (Sec.~\ref{ssec:pc-iii}). 

\subsection{Coverage Model}
\label{ssec:coverage_model}
The main lobe \gls{hpbw} for a UPA is given by~\cite{balanis2016antenna}
\begin{align}
\Theta _{P}(\theta, \phi)=\sqrt{\frac{1}{\cos^{2}\theta [\Theta _{L_x}^{-2}\cos^{2}\phi +\Theta _{L_y}^{-2}\sin^{2}\phi ]}},
\label{eq:Phi_P}
\end{align}
where $\theta$ and $\phi$ represent the antenna elevation and azimuth angles, respectively. In Eq.~\eqref{eq:Phi_P}, $\Theta _{L_x}$ ($\Theta _{L_y}$) is the \gls{hpbw} for a \gls{ula} of $N_{x}$ ($N_{y}$) elements along the x-axis (y-axis), i.e.,
\begin{align}\label{hpbw}
\Theta _{L_i}=\pi-2\cos^{-1}\left(\tfrac{1.391}{\pi \rho N_T}\right), \quad i\in \{x,\,y\},
\end{align} 
where $\rho=d/ \lambda$, $d$ is the antenna separation, $\lambda$ is wavelength, and $N_T=N_{x}\times N_{y}$ is the total number of antenna elements. 
For the VB, which is perpendicular to the ground (i.e., ${\theta=0}$), the HPBW can be calculated directly as 
\begin{equation}
\Theta _{P}^{\rm VB}=\Theta _{P}(0, \phi).
\label{eq:theta-vb}
\end{equation}

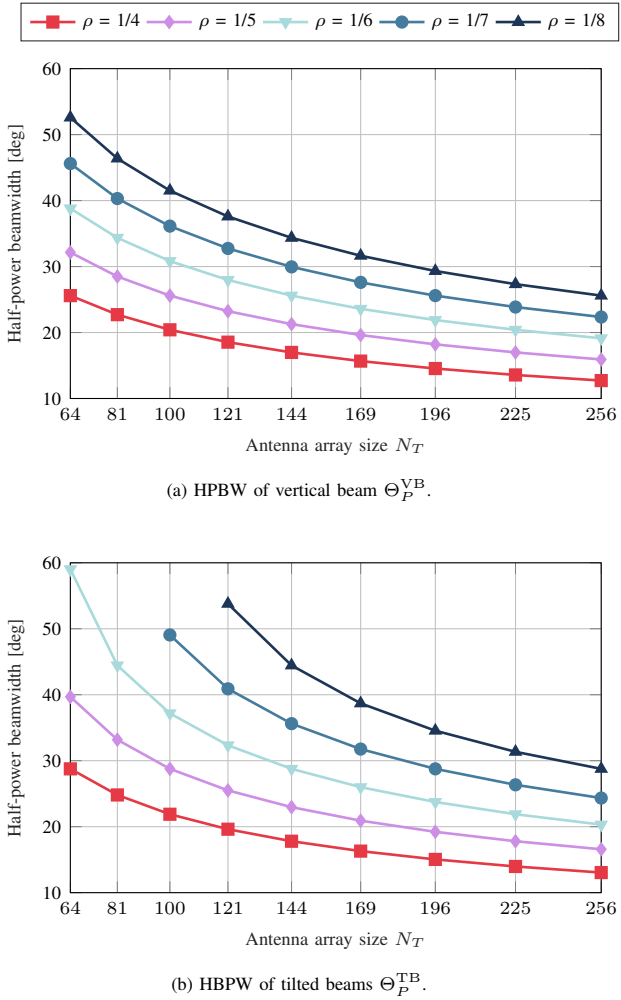
\begin{figure}[t!]
\centering
  \begin{subfigure}[t!]{0.44\textwidth}
  \centering
    \setlength\fwidth{0.93\textwidth}
    \setlength\fheight{0.55\textwidth}
%
%
\definecolor{mycolor1}{RGB}{230, 57, 70}%
\definecolor{mycolor2}{RGB}{205, 143, 229}%
\definecolor{mycolor3}{RGB}{168, 218, 220}%
\definecolor{mycolor4}{RGB}{69, 123, 157}%
\definecolor{mycolor5}{RGB}{29, 53, 87}%

%

\tikzset{
every pin/.append style={font=\scriptsize, align = left},
}
\begin{tikzpicture}

\pgfplotsset{
tick label style={font=\scriptsize},
label style={font=\scriptsize},
legend  style={font=\scriptsize}
}

\begin{axis}[%
width=0.951\fwidth,
height=\fheight,
at={(0\fwidth,0\fheight)},
scale only axis,
xmin=64,
xmax=256,
xtick={64,81,100,121,144,169,196,225,256},
xlabel style={font=\scriptsize\color{white!15!black}},
xticklabel style={align=center},
xlabel={Antenna array size $N_T$},
ymin=10,
ymax=60,
ytick={10,20,30,40,50,60},
yticklabels={10,20,30,40,50,60},
ylabel style={font=\scriptsize\color{white!15!black}},
ylabel={Half-power beamwidth [deg]},
axis background/.style={fill=white},
xmajorgrids,
ymajorgrids,
legend style={legend cell align=left, align=left, draw=white!15!black, at={(0.47,1.2)},/tikz/every even column/.append style={column sep=0.0cm},
  anchor=north ,legend columns=5}
]
\addplot [color=mycolor1, line width=1.0pt, mark size=2pt, mark=square*, mark options={solid, mycolor1}]
  table[row sep=crcr]{%
64	25.5937066671196\\
81	22.7096928296238\\
100	20.413012531184\\
121	18.5400925104353\\
144	16.9831516300967\\
169	15.668211831379\\
196	14.5427756643436\\
225	13.5685399928041\\
256	12.71689296256\\
};
\addlegendentry{ $\rho$ = 1/4}

\addplot [color=mycolor2, line width=1.0pt, mark size=2pt, mark=diamond*, mark options={solid, mycolor2}]
  table[row sep=crcr]{%
64	32.1466502301621\\
81	28.494289324463\\
100	25.5937066671196\\
121	23.2329214719562\\
144	21.2732465704482\\
169	19.6199810939768\\
196	18.2061818759255\\
225	16.9831516300967\\
256	15.9145998042111\\
};
\addlegendentry{ $\rho$ = 1/5}

\addplot [color=mycolor3, line width=1.0pt, mark size=2pt, mark=triangle*, mark options={solid, rotate=180, mycolor3}]
  table[row sep=crcr]{%
64	38.8094280916913\\
81	34.3540027307914\\
100	30.8281462202687\\
121	27.9655814853788\\
144	25.5937066671196\\
169	23.5954436219052\\
196	21.8884391572282\\
225	20.413012531184\\
256	19.12481989586\\
};
\addlegendentry{ $\rho$ = 1/6}

\addplot [color=mycolor4, line width=1.0pt, mark size=2pt, mark=*, mark options={solid, mycolor4}]
  table[row sep=crcr]{%
64	45.6116045715518\\
81	40.3078440015434\\
100	36.1293448042479\\
121	32.7474054425976\\
144	29.9514707911815\\
169	27.5999088374534\\
196	25.5937066671196\\
225	23.8614451567171\\
256	22.3502516367323\\
};
\addlegendentry{ $\rho$ = 1/7}

\addplot [color=mycolor5, line width=1.0pt, mark size=2pt, mark=triangle*, mark options={solid, mycolor5}]
  table[row sep=crcr]{%
64	52.588042788879\\
81	46.3774586793337\\
100	41.5117567050231\\
121	37.5885721060471\\
144	34.3540027307914\\
169	31.6390267147911\\
196	29.3263727295477\\
225	27.3319309739071\\
256	25.5937066671196\\
};
\addlegendentry{ $\rho$ = 1/8}

\end{axis}

\end{tikzpicture}%
    \caption{HPBW of vertical beam $\Theta _{P}^{\rm VB}$.}
      \label{fig:vb}
  \end{subfigure} \\\
  \begin{subfigure}[t!]{0.44\textwidth}
  \centering
    \setlength\fwidth{0.93\textwidth}
    \setlength\fheight{0.55\textwidth}
    \vspace{0.6cm}

\definecolor{mycolor1}{RGB}{230, 57, 70}%
\definecolor{mycolor2}{RGB}{205, 143, 229}%
\definecolor{mycolor3}{RGB}{168, 218, 220}%
\definecolor{mycolor4}{RGB}{69, 123, 157}%
\definecolor{mycolor5}{RGB}{29, 53, 87}%

\tikzset{
every pin/.append style={font=\scriptsize, align = left},
}
\begin{tikzpicture}

\pgfplotsset{
tick label style={font=\scriptsize},
label style={font=\scriptsize},
legend  style={font=\scriptsize}
}

\begin{axis}[%
width=0.951\fwidth,
height=\fheight,
at={(0\fwidth,0\fheight)},
scale only axis,
xmin=64,
xmax=256,
xtick={64,81,100,121,144,169,196,225,256},
xlabel style={font=\scriptsize\color{white!15!black}},
xticklabel style={align=center},
xlabel={Antenna array size $N_T$},
ymin=10,
ymax=60,
ytick={10,20,30,40,50,60},
yticklabels={10,20,30,40,50,60},
ylabel style={font=\scriptsize\color{white!15!black}},
ylabel={Half-power beamwidth [deg]},
axis background/.style={fill=white},
xmajorgrids,
ymajorgrids,
]
\addplot [color=mycolor1, line width=1.0pt, mark size=2pt, mark=square*, mark options={solid, mycolor1}]
  table[row sep=crcr]{%
64	28.7672567119567\\
81	24.8105211258619\\
100	21.8856868550496\\
121	19.6164886755969\\
144	17.7956747275761\\
169	16.2975774485517\\
196	15.0407201321059\\
225	13.9695795509206\\
256	13.0448252338696\\
};

\addplot [color=mycolor2, line width=1.0pt, mark size=2pt, mark=diamond*, mark options={solid, mycolor2}]
  table[row sep=crcr]{%
64	39.6836494446965\\
81	33.1805357544392\\
100	28.7672567119567\\
121	25.5025120178553\\
144	22.9611339037014\\
169	20.9134241432073\\
196	19.2213297177598\\
225	17.7956747275761\\
256	16.5757462107176\\
};

\addplot [color=mycolor3, line width=1.0pt, mark size=2pt, mark=triangle*, mark options={solid, rotate=180, mycolor3}]
  table[row sep=crcr]{%
64	59.0322961437118\\
81	44.4445849194308\\
100	37.1796183734211\\
121	32.3361381675532\\
144	28.7672567119567\\
169	25.9881942929999\\
196	23.7445558051527\\
225	21.8856868550496\\
256	20.3150350006798\\
};

\addplot [color=mycolor4, line width=1.0pt, mark size=2pt, mark=*, mark options={solid, mycolor4}]
  table[row sep=crcr]{%
100	49.0618407411216\\
121	40.8993958009644\\
144	35.6201357386683\\
169	31.7634700121195\\
196	28.7672567119567\\
225	26.3479470771952\\
256	24.3410363490338\\
};

\addplot [color=mycolor5, line width=1.0pt, mark size=2pt, mark=triangle*, mark options={solid, mycolor5}]
  table[row sep=crcr]{%
121	53.7846567214577\\
144	44.4445849194308\\
169	38.6947357489545\\
196	34.5520479822063\\
225	31.3493881364095\\
256	28.7672567119567\\
};

\end{axis}
\end{tikzpicture}%
    \caption{HBPW of tilted beams $\Theta _{P}^{\rm TB}$.} 
      \label{fig:tb}
  \end{subfigure}
  \setlength{\belowcaptionskip}{-0.66cm}
  \caption{Half-power beamwidth for different values of the antenna array size and antenna element separation distance.}
  \label{fig:hpbw} 
\end{figure}

As such, we define $r_{\rm VB}=h \tan(\Theta _{P}^{\rm VB}/2)$  the radius of the coverage area shaped by the VB, as depicted in Fig.~\ref{fig:scenario}.
For the TBs, which are adjacent to the VB, the HPBW is 
\begin{equation}
\Theta _{P}^{\rm TB}=\Theta _{P}\left((\Theta _{P}^{\rm VB}+\Theta _{P}^{\rm TB})/2, \phi\right).
\label{eq:theta-tb}
\end{equation}
In this case, we call $r_{\rm TB}=h \tan(\Theta _{P}^{\rm VB}/2+\Theta_{P}^{\rm TB})$  the distance between the projection of the UAV on the ground plane and the edge of the footprint shaped by the TB, as depicted in Fig.~\ref{fig:scenario}.
In Fig.~\ref{fig:hpbw}, we illustrate how the HPBW scales as a function of $\rho$ and $N_T$.
The results show that, while TBs exhibit a larger HPBW, especially when $\rho$ is small, the footprint of both VB and TBs is inversely proportional to the antenna size and grows with the antenna separation distance.
\medskip

\begin{theorem}\label{Theorem: coverage area}
According to the coverage model introduced above, the (circular) coverage area of a VB and the (oval) coverage area of a single TB are derived respectively~as
\begin{align}
&S_{\rm VB}=\pi r_{\rm VB}^{2}, \label{eq:area_vb}\\
&S_{\rm TB}=\pi ab,\label{eq:area_tb}
\end{align}
where parameters $a$ and $b$, which represent the two semi-axes of an ellipse, are given by
\begin{equation}
a=\tfrac{1}{2}\cdot h[\tan(\Theta_{H}^{\rm VB}+\Theta_{P}^{\rm TB})-\tan(\Theta_{H}^{\rm VB})],
\end{equation}
\vspace{0.13cm}
\medmuskip=2mu
\thickmuskip=2mu
\begin{align}
b=&\Big\{  (\sin\Theta_{H}^{\rm TB})^2-\Big[\cos\Theta_{H}^{\rm VB}\cdot \tan(\pi /2 -\Theta_{H}^{\rm VB}-\Theta_{H}^{\rm TB})-\notag\\
&-\arctan\left(\tfrac{h}{a+h \tan\Theta_{H}^{\rm VB}}\right)\Big]^2\Big\}^{1/2}/\notag\\
&\Big[{\cos(\Theta_{H}^{\rm VB}+\Theta_{P}^{\rm TB})\cdot \cos(\Theta _{H}^{\rm TB} /2)\cdot \sec(\Theta_{H}^{\rm VB}+\Theta_{H}^{\rm TB})}\Big] \cdot \notag\\
&\frac{\csc\left(\arctan\left(\tfrac{h}{a+h \tan\Theta_{H}^{\rm VB}}\right)\right)}{\sec\left(\pi /2 -\Theta_{H}^{\rm VB}-\Theta_{H}^{\rm TB}-\arctan\left(\tfrac{h}{a+h \tan\Theta_{H}^{\rm VB}}\right)\right)},  
\end{align}
\medmuskip=6mu
\thickmuskip=6mu
\vspace{0.13cm}

and $\Theta _{H}^{i}$ is the half of $\Theta _{P}^{i}$, $i\in\{\text{VB,TB}\}$. 

\textit{Proof:} It is obvious that the VB covers a circular area of radius $r_{\rm VB}=h \tan(\Theta _{H}^{\rm VB})$.
For the VBs,  it is common knowledge that the cross section of a cone and plane is an ellipse, which proves the assumption that TBs cover an oval area. The length of the two semi-axes of the ellipse can be derived by applying the well-known properties of similar triangles, and the details are omitted here.\hfill $\blacksquare$
\end{theorem}

Therefore, the coverage area of an ABF UAV, shaping one single VB, and that of an HBF UAV, forming $N_D$ parallel beams, can be expressed as
\begin{align}
&S_{\rm ABF}=S_{\rm VB},\label{eq:s_abf}\\
&S_{\rm HBF}=S_{\rm VB}+(N_{D}-1)S_{\rm TB}.
\label{eq:s_hbf}
\end{align}

\subsection{Power Consumption}
\label{ssec:pc-iii}
Finally, we can evaluate the power consumed by the UAVs to provide coverage service to an AoI of size $S_{\rm tot}$. Let  $N_{\rm ABF}=\ceil{{S_{\rm tot}}/{S_{\rm ABF}}}$ and $N_{\rm HBF}=\ceil{{S_{\rm tot}}/{S_{\rm HBF}}}$ be the number of  UAVs required to cover the AoI in case  ABF or HBF is implemented, respectively.  Overall, the total power consumption (which is due to both hovering and communication) for the ABF and HBF configurations is given based on Eqs.~\eqref{eq:p-tot-abf} and \eqref{eq:p-tot-hbf} by 
\begin{align}
&P_{C}^{\rm ABF}=N_{\rm ABF}\cdot P_{\rm tot}^{\rm ABF}=N_{\rm ABF}\cdot (P_{hover}+P_{\rm ABF}), \label{eq:P_C_ABF}\\
&P_{C}^{\rm HBF}=N_{\rm HBF}\cdot P_{\rm tot}^{\rm HBF}=N_{\rm HBF}\cdot (P_{hover}+P_{\rm HBF}).\label{eq:P_C_HBF}
\end{align}

\section{Stochastic Ergodic Capacity Analysis}
\label{sec:capacity}
In this section, we evaluate the stochastic ergodic capacity of a \gls{mmwave} UAV network implementing ABF or HBF. 
Specifically, in Sec.~\ref{ssec:association} we present the association rule for the \glspl{gu} and provide a probabilistic expression for the UAV-\gls{gu} serving distance, while in Secs.~\ref{ssec:ergodic-beams} and~\ref{ssec:ergodic-bf} we derive the  ergodic capacity experienced in the \gls{vb} or \glspl{tb} and when configuring ABF or HBF, respectively.

\subsection{Association Probabilities}
\label{ssec:association}
In order to facilitate the analysis, we evaluate the association probabilities between a reference UAV and \glspl{gu} from a \gls{2d} perspective.
Specifically, we make the assumption that UAVs form their beam(s) towards and communicate with the closest available \glspl{gu} within their respective coverage regions.
Indeed, let $r_{n}$ be the distance between the projection of the reference UAV on the user plane and the $n$-th closest user on the ground, which means that there are $(n-1)$ \glspl{gu} at distance smaller than $r_{n}$. Since the process $\Phi_u$ for the \glspl{gu}' distribution is a 2D homogeneous PPP with intensity measure $\lambda_u$, we can derive the \gls{cdf} $F_{r_n}(r)$ of $r_n$~as~\cite{6042301}
\medmuskip=4mu
\thickmuskip=4mu
\begin{align}
F_{r_n}(r)&=1-\mathbb{P}[r_{n} > r]\notag \\
&=1-\mathbb{P}[(n-1) \: \textrm{GUs closer than} \: r]  \notag\\
&=1-e^{- \lambda_{u} \pi r^{2}}\sum_{k=1}^{n-1}\frac{(\lambda_{u}  \pi r^{2})^{k}}{(k-1)!}.
\end{align}
\medmuskip=6mu
\thickmuskip=6mu
Consequently, the \gls{pdf} of $r_{n}$ can be derived as
\begin{align}
f_{r_{n}}(r)&=\frac{\mathrm{d}F_{r_{n}}(r)}{\mathrm{d}r} =r^{2n-1}(\lambda_{u} \pi)^{n}\frac{2}{(n-1)!}e^{- \lambda_{u} \pi r^{2}}.
\label{eq:f_r_n}
\end{align}

\begin{figure*}[!t]
\medmuskip=4mu
\thickmuskip=4mu
\begin{equation}
\label{eq_tau_vb}
\tau^{\rm VB} = B\int_{0}^{r_{\rm VB}} \int_{0}^{\infty} \int_{0}^{\infty} \frac{\left(l^{m_L-1}+\sum_{j=2}^{m_{L}}\frac{\Gamma(m_L)}{(m_L-j)!} l^{m_{L}-j}\right) e^{-\Delta_L}   }{(m_{L}-1)!} f_{r_1}^{\rm VB}(r)  p_\gamma(\gamma_N) \mathrm{d}\gamma_{N} \mathrm{d}t \mathrm{d}r
\tag{37}
\end{equation}
\begin{equation}
\label{eq_tau_tb}
\tau^{\rm TB}= B\int_{0}^{r_{\rm TB}} \int_{0}^{\infty} \int_{0}^{\infty} \frac{\left(l^{m_L-1}+\sum_{j=2}^{m_{L}}\frac{\Gamma(m_L)}{(m_L-j)!} l^{m_{L}-j}\right) e^{-\Delta_L}   }{(m_{L}-1)!} f_{r_{\tilde{n}}}^{\rm TB}(r)  p_\gamma(\gamma_N) \mathrm{d}\gamma_{N} \mathrm{d}t \mathrm{d}r
\tag{39}
\end{equation}
\rule[0.5ex]{\linewidth}{1pt}
\medmuskip=6mu
\thickmuskip=6mu
\end{figure*}

In case a single (vertical) beam is formed, i.e., ABF or HBF with $N_D=1$, the reference UAV steers the VB towards the closest  \gls{gu}. We can then evaluate the \gls{pdf} of the 2D distance between the UAV and the closest  \gls{gu}, i.e., $r_1$, from Eq.~\eqref{eq:f_r_n}~as 
\begin{equation}
    f_{r_1}^{\rm VB}(r)=e^{- \lambda_{u} \pi r^{2}}2\pi \lambda_{u} r.
    \label{eq:f_r_1}
\end{equation}
Since GUs' distribution is PPP with density $\lambda_u$, the average number of users within VB's footprint is equal to $\lambda_u S_{\rm VB}=\pi \lambda_u r_{\rm VB}^2$, where $r_{\rm VB}=h \tan(\Theta _{P}^{\rm VB}/2)$ is the radius of the coverage area described by the VB, as discussed in Sec.~\ref{ssec:coverage_model}.

In case tilted beams are also available, i.e., HBF with $2\leq N_D \leq N_{RF}$, the reference UAV forms a TB towards the $\tilde{n}$-th closest \gls{gu}, with $\widetilde{n}=\mathrm{floor}(\lambda_{u} \pi r_{\rm VB}^{2})+1$ corresponding to the first \gls{gu} outside  VB's coverage.
Based on the above definition, we can evaluate the \gls{pdf} of the 2D distance between the UAV and the $\tilde{n}$-th closest  \gls{gu}, i.e., $r_{\tilde{n}}$, from Eq.~\eqref{eq:f_r_n} as 
    \begin{equation}
        f_{r_{\tilde{n}}}^{\rm TB}(r)=r^{2{\tilde{n}}-1}(\lambda_{u} \pi)^{{\tilde{n}}}\frac{2}{({\tilde{n}}-1)!}e^{- \lambda_{u} \pi r^{2}}.
    \end{equation}

\subsection{Ergodic Capacity of Vertical and Tilted Beams}
\label{ssec:ergodic-beams}
In this subsection we provide an analytical closed-form expression for the ergodic capacity $\tau^{\rm VB}$ and $\tau^{\rm TB}$ for a target \gls{gu} within a VB or TB, respectively.
The ergodic capacity is a function of the \gls{snr} experienced by the GU attached to a reference UAV at distance $R=\sqrt{r^2+h^2}$ which, due to the presence of LOS and NLOS conditions for the channel, can be expressed as a function of $r$ as:
\begin{equation}
\text{SNR}(r)=\frac{P_{t}G\left[\gamma_{L}p_{L}(r)\ell_{L}(r)+\gamma_{N}p_{N}(r)\ell_{N}(r)\right]}{N_{D}\cdot  \mathrm{NF}\cdot \sigma ^{2}}.
\label{eq:SNR}
\end{equation}
In Eq.~\eqref{eq:SNR}, $P_{t}$ is the total transmitter power, $G$ is the beamforming gain, $\gamma_i$, $p_i(r)$, and $\ell_{i}(r)$, $i\in\{L,N\}$, represent, as introduced in Sec.~\ref{ssec:channel}, the small-scale fading, \gls{los} probability, and path gain, respectively, $\mathrm{NF}$ is the noise figure, and $\sigma ^{2}$ is the power of the thermal noise. Furthermore, $1\leq N_{D}\leq N_{RF}$ is the number of beams produced by the UAV simultaneously (where, of course, $N_{D}=1$ for ABF).\footnote{When HBF is used for transmission, the power available at each transmitting beam is given by the total transmit power divided by $N_D$, thus potentially reducing the received power~\cite{giordani2019tutorial}.}
\medskip

\setcounter{equation}{37}
 
\begin{theorem}\label{Theorem: rate for VB}
The ergodic capacity for a target \gls{gu} located within a VB can be written as in Eq.~\eqref{eq_tau_vb}, where $\Delta_L=m_L/\Omega_L\cdot \beta(t,\gamma_{N})$, $B$ is the system bandwidth, $p_\gamma(\gamma_N)$ is the probability distribution of the small-scale fading for NLOS~propagation, and
\medmuskip=2mu
\thickmuskip=2mu
\begin{align}
\beta(t,\gamma_{N})= \frac{(e^{t\ln2}-1)N_{D}\mathrm{NF}\sigma ^{2}-P_{t}G\gamma _{N}p_{N}(r)\ell_{N}(r)}{P_{t}Gp_{L}(r)\ell_{L}(r)}.
\end{align}
\medmuskip=6mu
\thickmuskip=6mu

\textit{Proof:} The proof is given in Appendix~\ref{app:th1}.\hfill $\blacksquare$
\end{theorem}
\medskip

\setcounter{equation}{39}
\begin{theorem}\label{Theorem: rate for TB}
The ergodic capacity for a target \gls{gu} located within a TB can be written as in Eq.~\eqref{eq_tau_tb}, where $r_{\rm TB}$ is the distance between the projection of the UAV on the ground plane and the farthest user within the TB, as described in Sec.~\ref{ssec:coverage_model}.

\textit{Proof:} The proof follows a similar method as that of Theorem~\ref{Theorem: rate for VB}, and is omitted here.\hfill $\blacksquare$
\end{theorem}

\begin{table*}[!t]
    \renewcommand{\arraystretch}{1.1}
    \setlength{\abovecaptionskip}{0.25cm}
    \caption{Parameters of the communication scenario and power consumption.}
    \label{communication scenario}
    \centering
    \begin{tabular}{lll|lll} 
        \hline
        \multicolumn{3}{c}{\emph{Communication scenario}} & \multicolumn{3}{c}{\emph{Power consumption}}\\
        \hline   
        Parameter & Description & Value & Parameter & Description & Value\\ 
        \hline      
        $S_{\rm tot}$ & Area of Interest          &1000 m$^{2}$ & $P_{PA}$ & Power amplifier~\cite{8333733} & $P_{t}/\xi , \xi =27\%$ \\     
        $\lambda_{u}$ & GU density      &\{0.005, 0.01, 0.05\} GUs/m$^{2}$ & $P_{PS}$ & Phase shifter~\cite{7313829}  & 21.6 mW \\ 
        $w$ & UAV weight~\cite{7991310}     &1.5  Kg & $b_{DAC}$ & DAC bits    & \{1,\dots,10\}   \\
        $c_1+c_2$ & UAV hovering parameters~\cite{7991310}  &    2.84 (m/kg)$^{1/2}$    & $P_{LO}$   & Local oscillator~\cite{4684642} & 22.5 mW \\            
        $h$ & UAV deployment height    &\{10,\dots,100\} m & $P_{C}$ & Combiner~\cite{yu201060} & 19.5 mW \\             
        $N_{x}\times N_{y}$ &Antenna array size  & $\{[9\times 9],\dots,[17\times 17]\}$& $P_{M}$ & Mixer~\cite{kraemer2011design} & 16.8 mW \\
        $\rho$ & Antenna separation~\cite{7913628} &$1/4$  &  $P_{H}$  & 90-deg hybrid coupler~\cite{Marcu:EECS-2011-132} & 3 mW \\
        $P_t$ & Transmit power  & 20 dBm & $P_{LPF}$ & Low pass filter~\cite{rangan2013energy}   & 14 mW \\
        $f_c$ & Carrier frequency & 28 GHz & $P_{BBamp}$ & Base-band amplifier~\cite{mendez2016hybrid}  & 5 mW \\
        $B$ & System bandwidth & 1 GHz & $P_{SP}$& Splitter~\cite{8057288} & 19.5 mW \\
         $\sigma^{2}$ & Thermal noise     & $-$84 dBm & $F_{S}$  & Sampling frequency    & 1 GHz  \\
       NF &  Noise figure     &5 dB & & \\
        \hline
    \end{tabular}
\end{table*}

\begin{table}[!t]
    \renewcommand{\arraystretch}{1.3}
    \setlength{\abovecaptionskip}{0.25cm}
    \caption{Parameters of the channel model.}
    \label{channel model}
    \centering
    \begin{tabular}{r|ll}
    \hline
    Parameter & LOS Value & NLOS Value \\\hline
    Nakagami-m shape parameter~\cite{8335329} & $\m_L = 3$ & $\m_N = 2$  \\
    Nakagami-m spread parameter~\cite{8335329} & $\Omega_L = 1$ & $\Omega_N = 1$  \\
    Additional attenuation factor~\cite{6834753} & $\eta_{L}=10^{-6.14}$ & $\eta_{N}=10^{-7.2}$\\
    Path loss exponent~\cite{6834753} & $\kappa_L = 2$ & $ \kappa_N = 2.92$ \\
    \hline
    LOS probability parameters~\cite{al2014optimal} & \multicolumn{2}{c}{$p=9.6117$ \: $q=0.1581$} \\ 
    \hline
    \end{tabular}
\end{table}

\subsection{Ergodic Capacity of ABF and HBF Architectures}
\label{ssec:ergodic-bf}
Based on Theorem~\ref{Theorem: rate for VB} and Theorem~\ref{Theorem: rate for TB}, we present the ergodic capacity experienced when implementing  an ABF ($\tau_{\rm ABF})$ or HBF ($\tau_{\rm HBF})$ architecture.
For ABF, the UAV steers a single VB, therefore 
\begin{align}
\tau_{\rm ABF}=\tau_{\rm VB}.
\label{eq:tau_ABF}
\end{align}
For HBF, the UAV steers one  VB and $N_D-1$ additional TBs, therefore
\begin{align}
\tau_{\rm HBF}=\frac{\tau_{\rm VB}+(N_{D}-1)\tau_{\rm TB}}{N_{D}}.
\label{eq:tau_HBF}
\end{align}

\section{Performance Evaluation}
\label{sec:evaluation}
In this section, we first introduce the simulation scenario and system parameters (Sec.~\ref{ssec:params}), then we present some numerical simulations to evaluate the power consumption (Sec.~\ref{ssec:results_power}) and  ergodic capacity (Sec.~\ref{ssec:results_capacity}) of UAV-assisted mmWave systems implementing ABF or HBF. The results are based on the analytical models described in Secs.~\ref{sec:power} and \ref{sec:capacity}, which are validated by Monte Carlo simulations.

\subsection{Evaluation Scenario and Parameters}
\label{ssec:params}
Table~\ref{communication scenario} summarizes the configuration parameters for the communication scenario. Notably, we consider a circular AoI of size $S_{\rm tot}=1000$ m$^{2}$, which represents a typical public safety scenario~\cite{navolio2011report}. GUs are then deployed according to a \gls{ppp} of density $\lambda_{u}\in\{0.005, 0.01, 0.05\}$ GUs/m$^{2}$. 
The reference UAV is an IRIS+ quadrotor of weight $w=1.5$ kg and parameters $c_{1}+c_{2}=2.84$ (m/kg)$^{1/2}$ measured in~\cite{7991310}, which hovers at height $h$ ranging from 10 to 100 m. UAVs are equipped with UPAs of $N_T=N_x\times N_y$ elements separated by a quarter of a wavelength to reduce grating lobes, i.e., $\rho=1/4$~\cite{7913628}, and operate at frequency $f_c=28$ GHz with a bandwidth $B=1$ GHz.
The transmit power $P_t$ and noise power $\sigma^2$ are set to $20$ dBm and $-84$ dBm, respectively, while the noise figure is equal to 5 dB~\cite{boschiero2020coverage}.

For the channel model, we consider the parameters in Table~\ref{channel model}: 
for an urban scenario, the S-curve parameters for the \gls{los} probability in Eq.~\eqref{eq:p_l} are given in~\cite{al2014optimal}, the path gain factors $\eta_{i}$ and $\kappa_{i}$, $i\in\{L,N\}$ are taken from~\cite{6834753}, while the  Nakagami fading values are adopted from~\cite{zhu2018secrecy}.

\begin{figure}[t!]
    \centering
    \begin{subfigure}[t!]{0.44\textwidth}
       \centering
    \setlength\fwidth{0.95\columnwidth}
    \setlength\fheight{0.7\columnwidth}
\definecolor{mycolor1}{RGB}{230, 57, 70}%
\definecolor{mycolor2}{RGB}{205, 143, 229}%
\definecolor{mycolor3}{RGB}{168, 218, 220}%
\definecolor{mycolor4}{RGB}{69, 123, 157}%
\definecolor{mycolor5}{RGB}{29, 53, 87}%
\tikzset{
every pin/.append style={font=\scriptsize, align = left},
}
\begin{tikzpicture}

\pgfplotsset{
tick label style={font=\scriptsize},
label style={font=\scriptsize},
legend  style={font=\scriptsize},
}
\pgfplotsset{scaled y ticks=false}

\begin{axis}[%
width=0.951\fwidth,
height=\fheight,
at={(0\fwidth,0\fheight)},
scale only axis,
xmin=81,
xmax=289,
xtick={ 81, 100, 121, 144, 169, 196, 225, 256, 289},
xlabel style={font=\scriptsize\color{white!15!black}},
ylabel style={font=\scriptsize\color{white!15!black}},
xlabel={Antenna array size $N_T$},
ymin=0,
ymax=70,
ylabel={Power consumption (W)},
axis background/.style={fill=white},
xmajorgrids,
ymajorgrids,
legend style={at={(0.124,0.607)}, anchor=south west, legend cell align=left, align=left, draw=white!15!black},
legend style={legend cell align=left, align=left, draw=white!15!black, at={(0.48,1.1)},/tikz/every even column/.append style={column sep=0.0cm},
  anchor=north ,legend columns=3}
]

\addplot [color=mycolor1, line width=1.0pt, mark size=3.3pt, mark=diamond*, mark options={solid, fill=mycolor1, mycolor1}]
  table[row sep=crcr]{%
81	2.31069037037037\\
100	2.72109037037037\\
121	3.17469037037037\\
144	3.67149037037037\\
169	4.21149037037037\\
196	4.79469037037037\\
225	5.42109037037037\\
256	6.09069037037037\\
289	6.80349037037037\\
};
\addlegendentry{HBF, $N_{RF}=1$}

\addplot [color=mycolor2, line width=1.0pt, mark size=3.3pt, mark=triangle*, mark options={solid, rotate=180, fill=mycolor2, mycolor2}]
  table[row sep=crcr]{%
81	7.77083037037037\\
100	9.37253037037037\\
121	11.1428303703704\\
144	13.0817303703704\\
169	15.1892303703704\\
196	17.4653303703704\\
225	19.9100303703704\\
256	22.5233303703704\\
289	25.3052303703704\\
};
\addlegendentry{HBF, $N_{RF}=3$}

\addplot [color=mycolor3, line width=1.0pt, mark size=3.3pt, mark=triangle*, mark options={solid, fill=mycolor3, mycolor3}]
  table[row sep=crcr]{%
81	11.6514703703704\\
100	14.0739703703704\\
121	16.7514703703704\\
144	19.6839703703704\\
169	22.8714703703704\\
196	26.3139703703704\\
225	30.0114703703704\\
256	33.9639703703704\\
289	38.1714703703704\\
};
\addlegendentry{HBF, $N_{RF}=5$}

\addplot [color=mycolor4, line width=1.0pt, mark size=3.3pt, mark=triangle*, mark options={solid, rotate=270, fill=mycolor4, mycolor4}]
  table[row sep=crcr]{%
81	15.5321103703704\\
100	18.7754103703704\\
121	22.3601103703704\\
144	26.2862103703704\\
169	30.5537103703704\\
196	35.1626103703704\\
225	40.1129103703704\\
256	45.4046103703704\\
289	51.0377103703704\\
};
\addlegendentry{HBF, $N_{RF}=7$}

\addplot [color=mycolor5, line width=1.0pt, mark size=3.3pt, mark=triangle*, mark options={solid, rotate=90, fill=mycolor5, mycolor5}]
  table[row sep=crcr]{%
81	19.4127503703704\\
100	23.4768503703704\\
121	27.9687503703704\\
144	32.8884503703704\\
169	38.2359503703704\\
196	44.0112503703704\\
225	50.2143503703704\\
256	56.8452503703704\\
289	63.9039503703704\\
};
\addlegendentry{HBF, $N_{RF}=9$}

\addplot [color=black, line width=1.0pt, dashed]
  table[row sep=crcr]{%
81	2.31069037037037\\
100	2.72109037037037\\
121	3.17469037037037\\
144	3.67149037037037\\
169	4.21149037037037\\
196	4.79469037037037\\
225	5.42109037037037\\
256	6.09069037037037\\
289	6.80349037037037\\
};
\addlegendentry{ABF}

\end{axis}
\end{tikzpicture}%
   \caption{Power consumption of the sole communication module.}
   \label{fig:pc-comm}
        \end{subfigure} \\\vspace{0.33cm}
    \begin{subfigure}[t!]{0.44\textwidth}
             \centering
    \setlength\fwidth{0.95\columnwidth}
    \setlength\fheight{0.7\columnwidth}
\definecolor{mycolor1}{RGB}{230, 57, 70}%
\definecolor{mycolor2}{RGB}{205, 143, 229}%
\definecolor{mycolor3}{RGB}{168, 218, 220}%
\definecolor{mycolor4}{RGB}{69, 123, 157}%
\definecolor{mycolor5}{RGB}{29, 53, 87}%
\tikzset{
every pin/.append style={font=\scriptsize, align = left},
}
\begin{tikzpicture}

\pgfplotsset{
tick label style={font=\scriptsize},
label style={font=\scriptsize},
legend  style={font=\scriptsize},
}
\pgfplotsset{scaled y ticks=false}

\begin{axis}[%
width=0.951\fwidth,
height=\fheight,
at={(0\fwidth,0\fheight)},
scale only axis,
xmin=81,
xmax=289,
xtick={ 81, 100, 121, 144, 169, 196, 225, 256, 289},
xlabel={Antenna array size $N_T$},
ymin=0,
ymax=60000,
ytick={10000,20000,30000,40000,50000,60000},
yticklabels={10,20,30,40,50,60},
ylabel style={font=\scriptsize\color{white!15!black}},
xlabel style={font=\scriptsize\color{white!15!black}},
ylabel={Power consumption [KW]},
axis background/.style={fill=white},
xmajorgrids,
ymajorgrids,
legend style={legend cell align=left, align=left, draw=white!15!black, at={(0.5,1.1)},/tikz/every even column/.append style={column sep=0.0cm},
  anchor=north ,legend columns=3}
]

\addplot [color=mycolor1, line width=1.0pt, mark size=3.3pt, mark=diamond*, mark options={solid, fill=mycolor1, mycolor1}]
  table[row sep=crcr]{%
81	13383.9824735045\\
100	16603.3079109618\\
121	20179.6537102568\\
144	24118.4630713893\\
169	28594.8981752783\\
196	33277.621060086\\
225	38512.7008876502\\
256	43967.5036961332\\
289	49991.6410473726\\
};
\addlegendentry{HBF, $N_D=1$}

\addplot [color=mycolor2, line width=1.0pt, mark size=3.3pt, mark=triangle*, mark options={solid, rotate=180, fill=mycolor2, mycolor2}]
  table[row sep=crcr]{%
81	5295.407628483\\
100	7052.44693767106\\
121	9013.03524777793\\
144	11010.4911578848\\
169	13401.6272698293\\
196	16022.7090826926\\
225	18884.0820964747\\
256	21996.8442111756\\
289	25188.9845258765\\
};
\addlegendentry{HBF, $N_D=2$}

\addplot [color=mycolor3, line width=1.0pt, mark size=3.3pt, mark=triangle*, mark options={solid, fill=mycolor3, mycolor3}]
  table[row sep=crcr]{%
81	3282.43849745732\\
100	4533.40214388896\\
121	5812.35339032061\\
144	7300.90365767106\\
169	8828.73772502151\\
196	10764.8108342096\\
225	12573.1402224788\\
256	14813.4812525857\\
289	17126.4888826926\\
};
\addlegendentry{HBF, $N_D=3$}

\addplot [color=mycolor4, line width=1.0pt, mark size=3.3pt, mark=triangle*, mark options={solid, rotate=270, fill=mycolor4, mycolor4}]
  table[row sep=crcr]{%
81	2445.80337286329\\
100	3357.53447745732\\
121	4294.46978205135\\
144	5441.15822756419\\
169	6808.71931399583\\
196	8035.78585950867\\
225	9687.44188685912\\
256	11400.7317142096\\
289	13180.73854156\\
};
\addlegendentry{HBF, $N_D=4$}

\addplot [color=mycolor5, line width=1.0pt, mark size=3.3pt, mark=triangle*, mark options={solid, rotate=90, fill=mycolor5, mycolor5}]
  table[row sep=crcr]{%
81	1943.04617010687\\
100	2685.94591378209\\
121	3453.07065745732\\
144	4432.15346205135\\
169	5447.95626664538\\
196	6695.60713215822\\
225	7800.02243675225\\
256	9350.7938631839\\
289	10767.5097286967\\
};
\addlegendentry{HBF, $N_D=5$}

\addplot [color=black, dashed, line width=1.0pt]
  table[row sep=crcr]{%
81	13383.9824735045\\
100	16603.3079109618\\
121	20179.6537102568\\
144	24118.4630713893\\
169	28594.8981752783\\
196	33277.621060086\\
225	38512.7008876502\\
256	43967.5036961332\\
289	49991.6410473726\\
};
\addlegendentry{ABF}

\end{axis}

\end{tikzpicture}%
   \caption{Total power consumption with $N_{RF}=N_D$.}
   \label{fig:pc}
        \end{subfigure}
    \caption{Power consumption vs. $N_T$ for ABF and HBF as a function of $N_{RF}$ and $N_D$, when  $h= 10$ m and $b_{DAC}$ = 6.}
    \label{rah_c}
\end{figure}
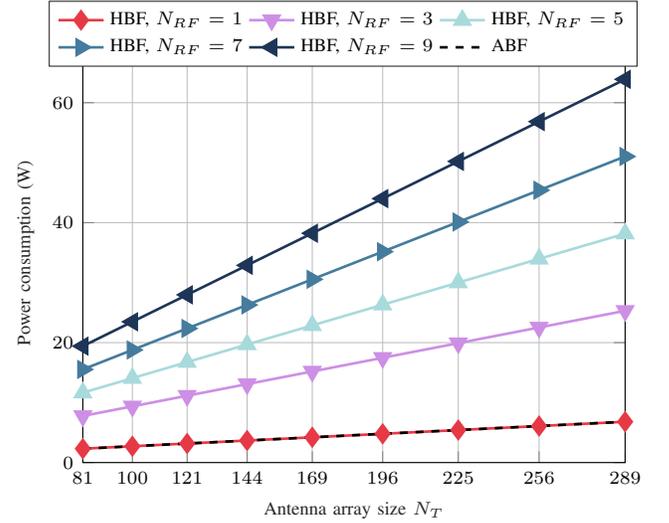
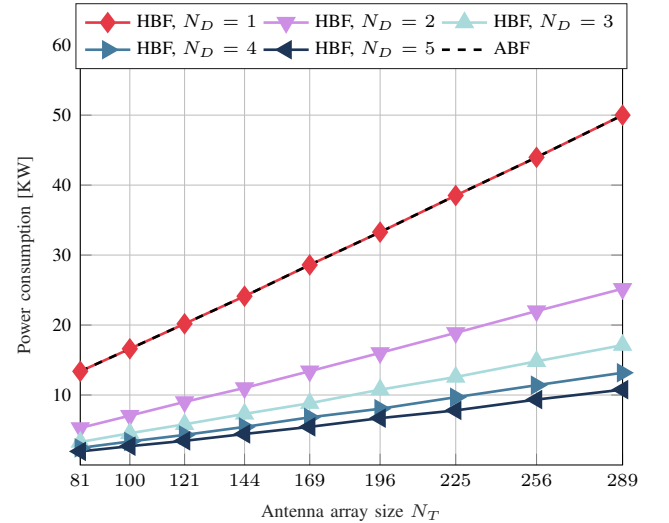



As described in Sec.~\ref{ssec:power}, a mmWave communication system implements several electronic components. 
In order to consider a relatively equitable power model for ABF and HBF architectures, we consider the most efficient implementation of each module and ideal parameter settings, as reported in Table~\ref{communication scenario}.
The power consumed by power amplifiers with \gls{pae} $\xi$ is given by $P_{PA}=P_t/\xi$, where where $P_{t}$ is the actual transmit power:
following~\cite{8333733}, we adopt $\xi=27\%$. 
We then implement active phase shifters which, at the expenses of lower linearity performance compared to passive elements, exhibit higher gain and resolution: the measured result shows that $P_{PS}=21.6$ mW~\cite{7313829}. 
Moreover, we consider the values reported in \cite{4684642} for local oscillators, \cite{yu201060} for combiners, \cite{kraemer2011design} for mixers, \cite{Marcu:EECS-2011-132} for $90$-deg hybrid couplers with buffers, \cite{rangan2013energy} for low pass filers,  \cite{mendez2016hybrid} for base-band amplifiers, and\cite{8057288} for splitters.
Finally, the number of DAC bits $b_{DAC}$ varies from 2 to 8, to consider different resolutions, while the sampling frequency is fixed to $F_{S}=1$ GHz.

\subsection{Evaluation Results: Power Consumption}
\label{ssec:results_power}


In this subsection we study the impact of the antenna array size, the number of RF chains and beams, and the DAC resolution, on the power consumption of UAV-assisted mmWave networks implementing an ABF or HBF configuration. 
In Fig.~\ref{fig:pc-comm} we plot the impact of the sole communication module on the UAV power consumption vs. the antenna array size.
As expected, the power consumption is a direct function of $N_T$. In fact, Eqs.~\eqref{eq:p_abp} and \eqref{eq:p_hbp} make it clear that configuring larger arrays would also proportionally increase the number of electronic components (specifically phase shifters) in both ABF and HBF architectures.
Still, ABF consumes less power than its HBF counterpart since it requires only a single RF chain.
The gap between the two schemes increases as $N_{RF}$ increases, due to the use of more power-consuming electronics, first and foremost DACs and phase splitters  in the hybrid~domain. 

\begin{figure}[t!] 
    \setlength\fwidth{0.9\columnwidth}
    \setlength\fheight{0.65\columnwidth}
\definecolor{mycolor1}{RGB}{230, 57, 70}%
\definecolor{mycolor2}{RGB}{205, 143, 229}%
\definecolor{mycolor3}{RGB}{168, 218, 220}%
\definecolor{mycolor4}{RGB}{69, 123, 157}%
\definecolor{mycolor5}{RGB}{29, 53, 87}%
\tikzset{
every pin/.append style={font=\scriptsize, align = left},
}
\begin{tikzpicture}

\pgfplotsset{
tick label style={font=\scriptsize},
label style={font=\scriptsize},
legend  style={font=\scriptsize},
}
\pgfplotsset{scaled y ticks=false}

\begin{axis}[%
width=0.951\fwidth,
height=\fheight,
at={(0\fwidth,0\fheight)},
scale only axis,
xmin=1,
xmax=5,
xtick={1,2,3,4,5},
xticklabels={10,20,30,40,50},
xlabel={UAV altitude [m]},
ymin=0,
ymax=140000,
ytick={20000,40000, 60000, 80000, 100000,120000,140000},
yticklabels={2,4,6, 8, 10, 12,14},
ylabel style={font=\scriptsize\color{white!15!black}},
xlabel style={font=\scriptsize\color{white!15!black}},
ylabel={Power consumption [KW]},
axis background/.style={fill=white},
xmajorgrids,
ymajorgrids,
legend style={legend cell align=left, align=left, draw=white!15!black, at={(0.5,1.1)},/tikz/every even column/.append style={column sep=0.0cm},
  anchor=north ,legend columns=3}
]

\addplot [color=mycolor1, line width=1.0pt, mark size=3.3pt, mark=diamond*, mark options={solid, fill=mycolor1, mycolor1}]
  table[row sep=crcr]{%
1	132334.126706776\\
2	33125.3566219236\\
3	14722.380720855\\
4	8364.98904594032\\
5	5353.5929894018\\
};
\addlegendentry{HBF, $N_D=1$}

\addplot [color=mycolor2, line width=1.0pt, mark size=3.3pt, mark=triangle*, mark options={solid, rotate=180, fill=mycolor2, mycolor2}]
  table[row sep=crcr]{%
1	52270.7978811547\\
2	13153.1092707481\\
3	5807.86643123941\\
4	3416.39201837613\\
5	2220.65481194448\\
};
\addlegendentry{HBF, $N_D=2$}

\addplot [color=mycolor3, line width=1.0pt, mark size=3.3pt, mark=triangle*, mark options={solid, fill=mycolor3, mycolor3}]
  table[row sep=crcr]{%
1	32824.3849745732\\
2	8292.4762041027\\
3	3800.71826021374\\
4	2073.11905102568\\
5	1382.07936735045\\
};
\addlegendentry{HBF, $N_D=3$}

\addplot [color=mycolor4, line width=1.0pt, mark size=3.3pt, mark=triangle*, mark options={solid, rotate=270, fill=mycolor4, mycolor4}]
  table[row sep=crcr]{%
1	24108.6332467953\\
2	6114.50843215822\\
3	2795.2038547009\\
4	1572.30216826926\\
5	1048.20144551284\\
};
\addlegendentry{HBF, $N_D=4$}

\addplot [color=mycolor5, line width=1.0pt, mark size=3.3pt, mark=triangle*, mark options={solid, rotate=90, fill=mycolor5, mycolor5}]
  table[row sep=crcr]{%
1	19077.1805792311\\
2	4769.29514480777\\
3	2119.68673102568\\
4	1236.48392643164\\
5	883.202804594031\\
};
\addlegendentry{HBF, $N_D=5$}

\addplot [color=black, dashed, line width=1.0pt]
  table[row sep=crcr]{%
1	132334.126706776\\
2	33125.3566219236\\
3	14722.380720855\\
4	8364.98904594032\\
5	5353.5929894018\\
};
\addlegendentry{ABF}

\end{axis}
\end{tikzpicture}%
   \caption{Total power consumption vs. $h$ for ABF and HBF as a function of $N_D$, when $N_T=9\times9$.}
   \label{fig:power-h}
  \end{figure}
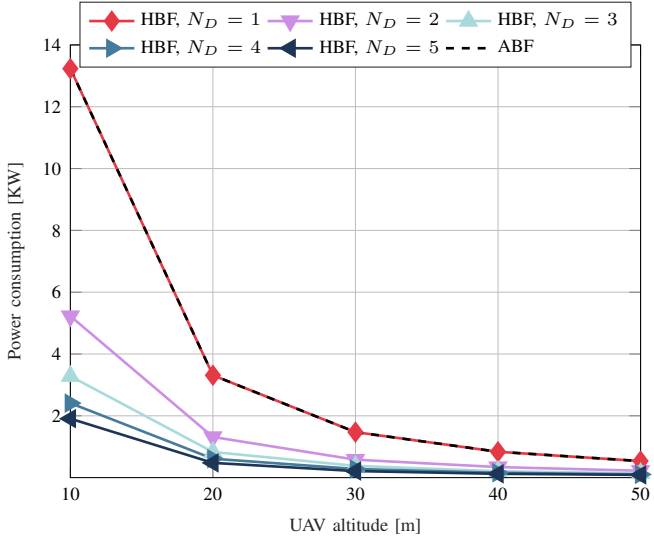
  
  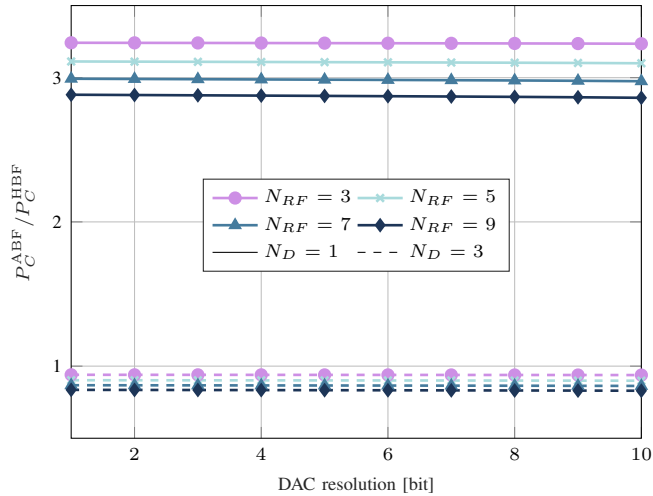
\begin{figure}[t!]
    \centering
    \setlength\fwidth{0.9\columnwidth}
    \setlength\fheight{0.65\columnwidth}
    \definecolor{mycolor1}{RGB}{230, 57, 70}%
\definecolor{mycolor2}{RGB}{205, 143, 229}%
\definecolor{mycolor3}{RGB}{168, 218, 220}%
\definecolor{mycolor4}{RGB}{69, 123, 157}%
\definecolor{mycolor5}{RGB}{29, 53, 87}%
\tikzset{
every pin/.append style={font=\scriptsize, align = left},
}
\begin{tikzpicture}

\pgfplotsset{
tick label style={font=\scriptsize},
label style={font=\scriptsize},
legend  style={font=\scriptsize},
}
\pgfplotsset{scaled y ticks=false}

\begin{axis}[%
width=0.951\fwidth,
height=\fheight,
at={(0\fwidth,0\fheight)},
scale only axis,
xmin=1,
xmax=10,
xlabel={DAC resolution [bit]},
ymin=0.5,
ymax=3.5,
ytick={1,2,3,4,5},
yticklabels={1,2,3,4,5},
ylabel={$P_C^{\rm ABF}/P_C^{\rm HBF}$},
ylabel style={font=\scriptsize\color{white!15!black}},
xlabel style={font=\scriptsize\color{white!15!black}},
axis background/.style={fill=white},
xmajorgrids,
ymajorgrids,
legend style={legend cell align=left, align=left, draw=white!15!black, at={(0.5,0.6)},/tikz/every even column/.append style={column sep=0.0cm},
  anchor=north ,legend columns=2}
]

\addplot [color=mycolor2, dashed, line width=1.0pt, mark size=2pt, mark=*, mark options={solid, mycolor2},forget plot]
  table[row sep=crcr]{%
1 0.940000802962354\\
2 0.939818152340546\\
3 0.939635005403349\\
4 0.939450757253635\\
5 0.939264198952687\\
6 0.939072914299457\\
7 0.938872074403837\\
8 0.938652029445707\\
9 0.938393499749222\\
10  0.938057982353532\\
};

\addplot [color=mycolor3, dashed, line width=1.0pt, mark size=2pt, mark=x, mark options={solid, mycolor3},forget plot]
  table[row sep=crcr]{%
1 0.902388383566291\\
2 0.902050098095066\\
3 0.90171101607055\\
4 0.901370019115243\\
5 0.901024873176765\\
6 0.90067111527435\\
7 0.900299829983147\\
8 0.899893210100671\\
9 0.899415700231594\\
10  0.898796355816543\\
};

\addplot [color=mycolor4, dashed, line width=1.0pt, mark size=2pt, mark=triangle*, mark options={solid, mycolor4},forget plot]
  table[row sep=crcr]{%
1 0.867670143709683\\
2 0.867200304995007\\
3 0.866729517113593\\
4 0.866256229098728\\
5 0.865777344358718\\
6 0.865286679559541\\
7 0.864771887955946\\
8 0.864208320370556\\
9 0.863546788516127\\
10  0.862689225115527\\
};

\addplot [color=mycolor5, dashed, line width=1.0pt, mark size=2pt, mark=diamond*, mark options={solid, mycolor5},forget plot]
  table[row sep=crcr]{%
1 0.835524409795543\\
2 0.834943122035114\\
3 0.834360839866308\\
4 0.833775646911972\\
5 0.833183719009644\\
6 0.832577423302901\\
7 0.83194152491163\\
8 0.831245622096784\\
9 0.830429079613617\\
10  0.829371098640641\\
};

 \addplot [color=mycolor2, line width=1.0pt, mark size=2pt, mark=*, mark options={solid, mycolor2}]
  table[row sep=crcr]{%
1 3.2420435857273\\
2 3.24141362746025\\
3 3.24078195741155\\
4 3.24014648930335\\
5 3.2395030535307\\
6 3.23884331666547\\
7 3.23815062396425\\
8 3.23739169339438\\
9 3.23650002974732\\
10  3.23534283709688\\
};
\addlegendentry{$N_{RF}=3$}

\addplot [color=mycolor3, line width=1.0pt, mark size=2pt, mark=x, mark options={solid, mycolor3}]
  table[row sep=crcr]{%
1 3.11231911883068\\
2 3.11115237914421\\
3 3.10998289216169\\
4 3.10880680062196\\
5 3.10761639932395\\
6 3.10639629553807\\
7 3.10511574014596\\
8 3.10371331646966\\
9 3.10206639467631\\
10  3.09993028842848\\
};
\addlegendentry{$N_{RF}=5$}

\addplot [color=mycolor4, line width=1.0pt, mark size=2pt, mark=triangle*, mark options={solid, mycolor4}]
  table[row sep=crcr]{%
1 2.99257661810074\\
2 2.99095615396237\\
3 2.98933241616729\\
4 2.98770005546296\\
5 2.98604839176782\\
6 2.98435609888903\\
7 2.98258059315418\\
8 2.98063686005355\\
9 2.97835525018828\\
10  2.9753975315209\\
};
\addlegendentry{$N_{RF}=7$}

\addplot [color=mycolor5, line width=1.0pt, mark size=2pt, mark=diamond*, mark options={solid, mycolor5}]
  table[row sep=crcr]{%
1 2.88170663786626\\
2 2.87970178824356\\
3 2.87769350892666\\
4 2.87567519036986\\
5 2.87363364311489\\
6 2.87154254159572\\
7 2.86934934102174\\
8 2.86694918641544\\
9 2.86413294805513\\
10  2.86048399327078\\
};
\addlegendentry{$N_{RF}=9$}


\addplot [color=black]
  table[row sep=crcr]{%
-1 -1\\
};
\addlegendentry{$N_D=1$}

\addplot [color=black, dashed]
  table[row sep=crcr]{%
-1 -1\\
};
\addlegendentry{$N_D=3$}


\end{axis}

\end{tikzpicture}%
\caption{$P_C^{\rm ABF}/P_C^{\rm HBF}$ vs. the DAC resolution and the number of RF chains, when $N_D=1$ and $N_D=3$.}
   \label{fig:bdac}
  \end{figure}

Nevertheless, Fig.~\ref{fig:pc} demonstrates that, if we investigate the total power consumption due to both hovering and communication for the ABF and HBF schemes, results would be radically different. 
Notably, despite using more power-hungry blocks, HBF results in more efficient operations than ABF when $N_{D}>1$. In particular, power consumption can be reduced by up to 4.5 times when HBF with $N_D=5$ is adopted.
This is due to the fact that the total power consumption is dominated by the hovering power, which is proportional to the UAV swarm size, since the communication module, which consumes up to 63 W when $N_T = 289$ (as illustrated in  Fig.~\ref{fig:pc-comm}) accounts only for a small amount of the overall power budget. 
Indeed, from  Eqs.~\eqref{eq:P_C_ABF} and \eqref{eq:P_C_HBF}, it is clear that HBF, which forms  multiple parallel TBs simultaneously,  can cover a larger service area compared to ABF, thus potentially reducing the number of UAVs that need to be deployed to achieve full service coverage, i.e., ${N_{\rm HBF}\leq N_{\rm ABF}}$.
According to Eq.~\eqref{eq:theta-tb}, $N_{\rm HBF}$ can be also reduced by increasing $h$, thereby geometrically extending the coverage region shaped by the TBs projected on the ground, which would result in even lower power consumption with respect to ABF, as illustrated in Fig.~\ref{fig:power-h}.

Finally, Fig.~\ref{fig:bdac} shows the effect of the DAC resolution and the number of RF chains on the total power consumption. 
We observe that the ratio between the power consumed when ABF is used  ($P_C^{\rm ABF}$) and that when HBF is preferred ($P_C^{\rm HBF}$) is greater than 1 when $N_D>1$, thus acknowledging the conclusion that HBF results in more efficient UAV operations than ABF when multiple parallel TBs are steered.
The optimal deployment is to set $N_D=N_{RF}$, so as to avoid using more RF chains (the most power consuming electronic block) than the actual number of beams used for communication. 
Finally, it is evident that both $P_C^{\rm ABF}$ and $P_C^{\rm HBF}$ are almost independent of the DAC resolution.
Such commonly accepted conclusion that HBF suffers from high power consumption is the result of implicitly assuming that high resolution and wide band DACs dominate the overall energy budget~\cite{giordani2017improved} while, in turn, the main power constraint is represented by the hovering power, i.e., the UAV swarm size, rather than communication hardware components.

\subsection{Evaluation Results: Ergodic Capacity}
\label{ssec:results_capacity}
We now investigate the downlink ergodic capacity for ABF and HBF architectures as a function of different UAV-specific parameter configurations. 
In the following figures, the markers indicate the Monte Carlo simulation results, which we obtained by generating 1000 random realizations of a PPP for the GUs' distribution across the AoI of size $S_{\rm tot}=1000$ m$^2$, while the lines represent the numerical results for the analytical model, solved using the MATLAB Symbolic toolbox.

\begin{figure}[t!]
    \centering
    \begin{subfigure}[t!]{0.44\textwidth}
     \setlength\fwidth{0.94\columnwidth}
    \setlength\fheight{0.7\columnwidth}
%
%
\definecolor{mycolor1}{RGB}{230, 57, 70}%
\definecolor{mycolor2}{RGB}{205, 143, 229}%
\definecolor{mycolor3}{RGB}{168, 218, 220}%
\definecolor{mycolor4}{RGB}{69, 123, 157}%
\definecolor{mycolor5}{RGB}{29, 53, 87}%
\tikzset{
every pin/.append style={font=\scriptsize, align = left},
}
\begin{tikzpicture}

\pgfplotsset{
tick label style={font=\scriptsize},
label style={font=\scriptsize},
legend  style={font=\scriptsize},
}
\pgfplotsset{scaled y ticks=false}

\begin{axis}[%
width=0.951\fwidth,
height=\fheight,
at={(0\fwidth,0\fheight)},
scale only axis,
xmin=10,
xmax=100,
xlabel style={font=\color{white!15!black}},
xlabel={UAV altitude [m]},
ymin=1000000000,
ymax=10000000000,
xlabel style={font=\scriptsize\color{white!15!black}},
ylabel={Ergodic capacity $\tau_{\rm ABF}$ [Gbps]},
ytick={2000000000,4000000000,6000000000,8000000000,10000000000},
yticklabels={2, 4, 6, 8, 10},
axis background/.style={fill=white},
ylabel style={font=\scriptsize\color{white!15!black}},
xmajorgrids,
ymajorgrids,
legend style={legend cell align=left, align=left, draw=white!15!black, at={(0.47,1.25)},/tikz/every even column/.append style={column sep=0.0cm},
  anchor=north ,legend columns=3}
]
\addplot [color=mycolor1, line width=1.0pt]
  table[row sep=crcr]{%
10	5583313100.59167\\
20	9127749585.19066\\
30	8727574739.55412\\
40	7934093141.10079\\
50	7296738507.96429\\
60	6776806626.42999\\
70	6338584359.53541\\
80	5960472168.91845\\
90	5628494743.85118\\
100	5333086123.26751\\
};
\addlegendentry{$N_T=9\times9$}

\addplot [color=mycolor2, line width=1.0pt]
  table[row sep=crcr]{%
10	4268831001.38834\\
20	8526817203.63613\\
30	9118498807.33565\\
40	8500427440.61954\\
50	7871728286.6225\\
60	7350382402.89646\\
70	6910248663.00233\\
80	6529956556.84743\\
90	6195542979.07132\\
100	5897453450.06114\\
};
\addlegendentry{$N_T=11\times11$}

\addplot [color=mycolor3, line width=1.0pt]
  table[row sep=crcr]{%
10	3336135799.51903\\
20	7636318911.46984\\
30	9140174991.55004\\
40	8914252903.09491\\
50	8346822654.36684\\
60	7829080427.92901\\
70	7388003687.36016\\
80	7006440838.86445\\
90	6670597791.8888\\
100	6370928275.17513\\
};
\addlegendentry{$N_T=13\times13$}

\addplot [color=mycolor4, line width=1.0pt]
  table[row sep=crcr]{%
10	2667995202.64357\\
20	6709839849.38468\\
30	8844811685.23209\\
40	9134945717.7941\\
50	8729380127.5337\\
60	8237408664.26656\\
70	7798079483.72711\\
80	7415846995.32657\\
90	7079100733.50048\\
100	6778424889.55125\\
};
\addlegendentry{$N_T=15\times15$}

\addplot [color=mycolor5, line width=1.0pt]
  table[row sep=crcr]{%
10	2178573037.05161\\
20	5858809169.0883\\
30	8347519715.78438\\
40	9145769246.25967\\
50	9001049160.60749\\
60	8582664374.22456\\
70	8155822480.78186\\
80	7774529197.91554\\
90	7437285394.92174\\
100	7135936276.99778\\
};
\addlegendentry{$N_T=17\times17$}

\addplot [color=mycolor1, only marks, mark size=3.3pt, mark=triangle*, mark options={solid, fill=mycolor1, mycolor1}, forget plot]
  table[row sep=crcr]{%
10	5692274449.46832\\
20	9128148220.26756\\
30	8668360912.45302\\
40	7904966875.80501\\
50	7231004298.09684\\
60	6779243909.58334\\
70	6311013266.27872\\
80	5917639995.79968\\
90	5646741793.32276\\
100	5320837605.36025\\
};

\addplot [color=mycolor2, only marks, mark size=3.3pt, mark=triangle*, mark options={solid, fill=mycolor2, mycolor2}, forget plot]
  table[row sep=crcr]{%
10	4018680086.07866\\
20	8653227775.60355\\
30	9110772048.05864\\
40	8501947886.40238\\
50	7894994872.28884\\
60	7351710940.01963\\
70	6911039497.31998\\
80	6510375122.28809\\
90	6203173619.3967\\
100	5832813805.32396\\
};

\addplot [color=mycolor3, only marks, mark size=3.3pt, mark=triangle*, mark options={solid, fill=mycolor3, mycolor3}, forget plot]
  table[row sep=crcr]{%
10	3562335551.99531\\
20	7448877592.6783\\
30	9319193703.12702\\
40	8919255412.17241\\
50	8361635965.55494\\
60	7880828508.47162\\
70	7382356307.5089\\
80	6946069103.10219\\
90	6687843822.85745\\
100	6336722972.6018\\
};

\addplot [color=mycolor4, only marks, mark size=3.3pt, mark=triangle*, mark options={solid, fill=mycolor4, mycolor4}, forget plot]
  table[row sep=crcr]{%
10	2612252070.67916\\
20	6467842298.64768\\
30	8721495995.7624\\
40	9202668007.02529\\
50	8766622611.30641\\
60	8247627620.03197\\
70	7838146369.28539\\
80	7445399860.78155\\
90	7139502345.59647\\
100	6755037372.13839\\
};

\addplot [color=mycolor5, only marks, mark size=3.3pt, mark=triangle*, mark options={solid, fill=mycolor5, mycolor5}, forget plot]
  table[row sep=crcr]{%
10	2342991493.55271\\
20	6162734005.91841\\
30	8229930872.2159\\
40	9117722217.93696\\
50	8995528746.13029\\
60	8590024550.3741\\
70	8130823902.43921\\
80	7732702835.53023\\
90	7409500582.86891\\
100	7163597154.89076\\
};


\addplot [color=black]
  table[row sep=crcr]{%
-1 -1\\
};
\addlegendentry{Analysis}

\addplot [color=black, draw=none, mark size=1.3pt, mark=triangle*, mark options={solid, fill=black, black}, only marks]
  table[row sep=crcr]{%
-1 -1\\
};
\addlegendentry{Monte Carlo}

\end{axis}

\end{tikzpicture}%
    \caption{Analog beamforming architecture (VB only).}
        \label{ra_c}
        \end{subfigure} \\
    \begin{subfigure}[t!]{0.44\textwidth}
         \setlength\fwidth{0.94\columnwidth}
    \setlength\fheight{0.7\columnwidth}
    \vspace{0.33cm}
    %
%
\definecolor{mycolor1}{RGB}{230, 57, 70}%
\definecolor{mycolor2}{RGB}{205, 143, 229}%
\definecolor{mycolor3}{RGB}{168, 218, 220}%
\definecolor{mycolor4}{RGB}{69, 123, 157}%
\definecolor{mycolor5}{RGB}{29, 53, 87}%
\tikzset{
every pin/.append style={font=\scriptsize, align = left},
}
\begin{tikzpicture}

\pgfplotsset{
tick label style={font=\scriptsize},
label style={font=\scriptsize},
legend  style={font=\scriptsize},
}
\pgfplotsset{scaled y ticks=false}

\begin{axis}[%
width=0.951\fwidth,
height=\fheight,
at={(0\fwidth,0\fheight)},
scale only axis,
xmin=10,
xmax=100,
xlabel style={font=\color{white!15!black}},
xlabel={UAV altitude [m]},
ymin=1000000000,
ymax=10000000000,
xlabel style={font=\scriptsize\color{white!15!black}},
ylabel={Ergodic capacity $\tau_{\rm HBF}$ [Gbps]},
ytick={2000000000,4000000000,6000000000,8000000000,10000000000},
yticklabels={2, 4, 6, 8, 10},
axis background/.style={fill=white},
ylabel style={font=\scriptsize\color{white!15!black}},
xmajorgrids,
ymajorgrids,
]
\addplot [color=mycolor1, line width=1.0pt]
  table[row sep=crcr]{%
10	7979460614.41732\\
20	8539294021.0841\\
30	7723285864.17918\\
40	6913825620.8442\\
50	6281512233.20486\\
60	5766095086.65699\\
70	5332761394.98535\\
80	4961555841.53581\\
90	4636672636.84198\\
100	4349197146.72301\\
};

\addplot [color=mycolor2, line width=1.0pt]
  table[row sep=crcr]{%
10	7629679895.5202\\
20	8592470399.66242\\
30	8225874265.99701\\
40	7494403885.11504\\
50	6861588032.45004\\
60	6342739771.25767\\
70	5906910372.90983\\
80	5531175270.62576\\
90	5201700927.68281\\
100	4908978547.28041\\
};

\addplot [color=mycolor3, line width=1.0pt]
  table[row sep=crcr]{%
10	7243162650.94324\\
20	8445903077.45888\\
30	8504331280.76266\\
40	7949458235.94187\\
50	7342469270.47165\\
60	6824563472.99319\\
70	6385865436.16406\\
80	6007040710.84947\\
90	5674848653.9617\\
100	5379454174.79883\\
};

\addplot [color=mycolor4, line width=1.0pt]
  table[row sep=crcr]{%
10	6756875570.61701\\
20	8252833238.64175\\
30	8600565607.04288\\
40	8277712387.80928\\
50	7743355322.93913\\
60	7236109159.95903\\
70	6797009341.07058\\
80	6416485686.93172\\
90	6082880332.57269\\
100	5784794223.46413\\
};

\addplot [color=mycolor5, line width=1.0pt]
  table[row sep=crcr]{%
10	6205581625.01959\\
20	8047653356.1411\\
30	8570187328.93247\\
40	8483004281.4213\\
50	8064951747.20355\\
60	7589414623.26188\\
70	7155701953.4741\\
80	6775023608.52488\\
90	6440377364.09166\\
100	6140986016.32855\\
};

\addplot [color=mycolor1, only marks, mark size=3.3pt, mark=triangle*, mark options={solid, fill=mycolor1, mycolor1},forget plot]
  table[row sep=crcr]{%
10	8011674183.58371\\
20	8515131450.86399\\
30	7672869204.68537\\
40	6884031433.30673\\
50	6214744259.0854\\
60	5767184913.56321\\
70	5306102672.31497\\
80	4919479223.47458\\
90	4654559660.12758\\
100	4336932080.14847\\
};

\addplot [color=mycolor2, only marks, mark size=3.3pt, mark=triangle*, mark options={solid, fill=mycolor2, mycolor2},forget plot]
  table[row sep=crcr]{%
10	7425755281.36873\\
20	8649554443.36427\\
30	8205794459.85337\\
40	7492671189.74978\\
50	6883934947.31738\\
60	6343888147.00869\\
70	5907222316.52554\\
80	5511126613.26575\\
90	5208541276.22788\\
100	4844972625.29811\\
};

\addplot [color=mycolor3, only marks, mark size=3.3pt, mark=triangle*, mark options={solid, fill=mycolor3, mycolor3},forget plot]
  table[row sep=crcr]{%
10	7221291678.85561\\
20	8339425185.04004\\
30	8592097763.13146\\
40	7924670156.72573\\
50	7353573038.3044\\
60	6874992798.15514\\
70	6379372360.36537\\
80	5947398229.66426\\
90	5691629662.85921\\
100	5345512456.59978\\
};

\addplot [color=mycolor4, only marks, mark size=3.3pt, mark=triangle*, mark options={solid, fill=mycolor4, mycolor4},forget plot]
  table[row sep=crcr]{%
10	6560617984.9922\\
20	8126403325.37964\\
30	8518271951.54766\\
40	8300265837.02683\\
50	7765368146.03418\\
60	7243412445.4473\\
70	6835803597.3485\\
80	6445438669.02943\\
90	6142186026.65426\\
100	5761300057.83288\\
};

\addplot [color=mycolor5, only marks, mark size=3.3pt, mark=triangle*, mark options={solid, fill=mycolor5, mycolor5},forget plot]
  table[row sep=crcr]{%
10	6100269261.01541\\
20	8162558108.51953\\
30	8529329922.66947\\
40	8503219377.20039\\
50	8085219802.00396\\
60	7586807880.86917\\
70	7134480998.18126\\
80	6733463031.4531\\
90	6412430069.52898\\
100	6168290225.29405\\
};
\addplot [draw=none, color=white, only marks]
  table[row sep=crcr]{%
-1 -1\\
};

\addplot [color=black]
  table[row sep=crcr]{%
-1 -1\\
};

\addplot [color=black, draw=none, mark size=1.3pt, mark=triangle*, mark options={solid, fill=black, black}, only marks]
  table[row sep=crcr]{%
-1 -1\\
};

\end{axis}

\end{tikzpicture}%
    \caption{Hybrid beamforming architecture (VB and TBs).}
        \label{rh_c}
        \end{subfigure}
    \caption{Ergodic capacity of ABF and HBF vs. the deployment altitude and the antenna array size, when $N_{D}=2$ (for HBF) and $\lambda_{u}=0.05$ GU/m$^2$. The lines represent the results from the analytical model, and the markers indicate Monte Carlo simulations.}
    \label{rah_c}
\end{figure}
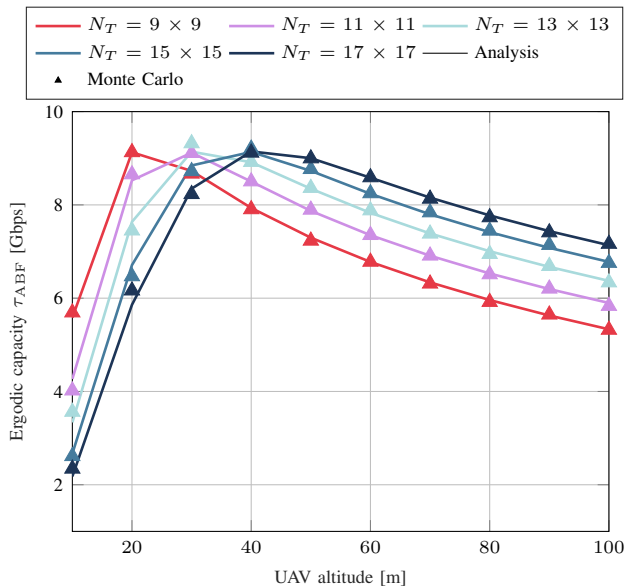
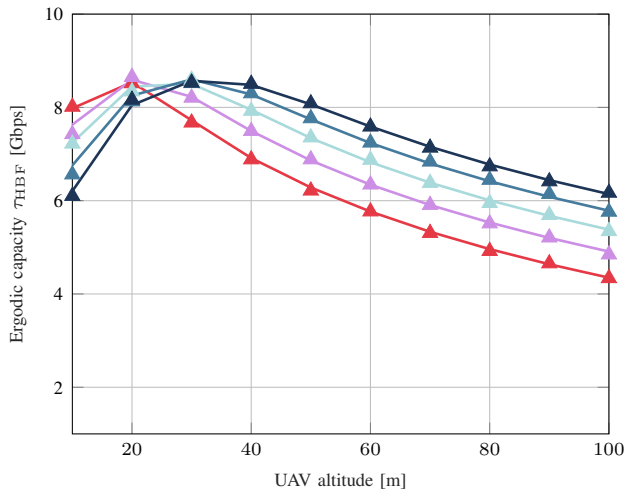

Figs. \ref{ra_c} and \ref{rh_c} show the downlink ergodic capacity when UAVs implement ABF or HBF with $N_D=2$, respectively, vs. the deployment altitude  $h$ and the antenna array size, with $\lambda_{u}=0.05$ GU/m$^2$.
First, it is clear that the numerical results closely follow the analytic curves representing
Eqs.~\eqref{eq:tau_ABF} and~\eqref{eq:tau_HBF}, thereby validating our theoretical framework. 
Second, we see that the large spectrum available at \glspl{mmwave} makes it possible to achieve multi-Gbps transmission rates, thereby opening up new opportunities for next-generation public safety networks~\cite{tracy20155g}.
Third, we observe that the curves of the ergodic capacity are bell shaped, and the peak value appears at a specific altitude $h^*$, which typically ranges between $20$ and $40$ m.
On one hand, when the UAVs are deployed at low altitudes, the LOS probability is small, and the increased path loss in NLOS results in degraded channels. Consequently, $\tau_{\rm ABF}$ and $\tau_{\rm HBF}$  improve for increasing $h$, thus establishing more favorable propagation conditions.
On the other hand, for higher altitudes, even if the link is likely in LOS, the impact of the increased distance between the GUs and their serving UAVs decreases the overall link budget, preventing successful communications.
As expected, the ABF capacity (Fig.~\ref{ra_c}) for $h=h^*$ is  higher than its HBF counterpart (Fig.~\ref{rh_c}). This is due to the fact that, in HBF, the transmit power has to be split among $N_D$ parallel streams, thus reducing the received power in the  VB and TBs and, consequently, the achievable data rate. 
However, while HBF's capacity degradation is quite limited (e.g., $-7\%$ at $h^*=30$~m when $N_T=13\times13$), the hybrid approach has the potential to improve the ergodic capacity when $h<h^*$ compared to an analog architecture. 
At low altitudes, in fact, according to the HPBW model in Eq.~\eqref{eq:theta-vb}, the (single) analog VB shapes a very small coverage region on the ground plane, increasing the risk of leaving some GUs unserved, which may instead lay within the coverage umbrella of HBF's TBs: at $h=10$ m, HBF improves the capacity by up to $33\%$   compared to ABF when $N_T=17\times17$.

Fig.~\ref{rah_c} also exemplifies the significant impact of the antenna array size on the overall capacity performance.
If $h<h^*$, it becomes convenient to reduce the number of antenna elements, which also results in lower power consumption for the UAV (see Fig.~\ref{fig:pc}). 
In fact,  reducing $N_T$ permits to produce wider beams (as also depicted in Fig.~\ref{fig:hpbw}), thus increasing the coverage area shaped by the projection of the VB and TBs onto the ground which, at low altitude, is quite limited: at $h=10$ m, $r_{\rm VB}$  doubles from $1.047$ m when $N_T=17\times17$  to $2$ m when $N_T=9\times9$.
If $h>h^*$, the beamwidth is already sufficiently large to allow for continuous coverage (thanks to the widening of the beam's projection on the ground with the distance), while increasing the array size permits to achieve higher gains by beamforming and combat the severe path loss experienced at high altitudes.



\begin{figure}[t!]
    \centering
    \setlength{\belowcaptionskip}{-0.66cm}
    \setlength\fwidth{0.5\columnwidth}
    \setlength\fheight{0.4\columnwidth}
    \includegraphics[width=0.99\columnwidth]{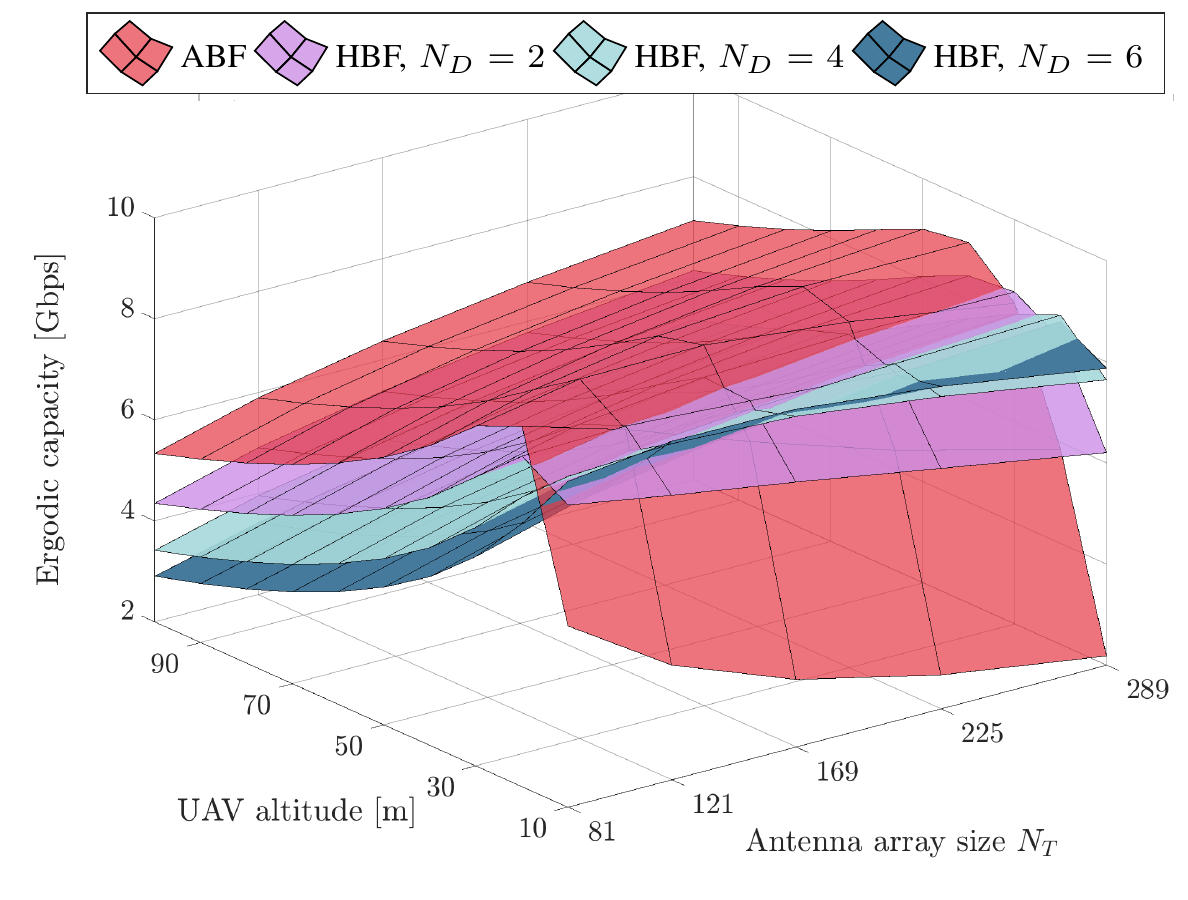}
    \caption{Ergodic capacity of ABF and HBF vs. the deployment altitude, the antenna array size, and the number of parallel beams in HBF, when $\lambda_{u}=0.05$ GU/m$^2$.}
    \label{3d0.05}
\end{figure}

The same conclusions can be derived  from Fig.~\ref{3d0.05}, which shows the ergodic capacity of the ABF and HBF schemes as a function of $N_{D}$.
We can see that HBF indeed exhibits higher capacity than ABF at lower altitudes, especially when considering large-size antenna arrays. In these circumstances, the performance improvement grows consistently with $N_D$, i.e., as  more parallel hybrid TBs are configured; e.g., when $N_T = 17 \times 17$ and $h = 10$ m, HBF with $N_{D} = 6$ guarantees the highest available capacity.
However, for larger values of $h$, HBF underperforms ABF since the transmit power has to be subdivided among $N_D$ parallel beams, thus further reducing the received power which is already significantly deteriorated by the strong path loss experienced at  high altitude.

The GU density is another key parameter of the network.
Along these lines, Fig.~\ref{3D} illustrates the ergodic capacity of the ABF and HBF architectures for different values of $\lambda_{u}$. 
We clearly see that ABF's capacity drops when $\lambda_{u}$ decreases.
The reasons are twofold. 
First, considering sparser networks,  Eq.~\eqref{eq:f_r_1} demonstrates that the distance to the serving UAV increases, which results in  weaker channels.
Second, when fewer GUs are deployed, the probability that they will be placed within the coverage of the single VB decreases accordingly. In  this perspective, an HBF configuration has the potential to guarantee better capacity by making it possible for the (sparse) GUs to still maintain connectivity with one of the available TBs:
the performance gap grows proportionally with $N_D$, since more parallel TBs result in more connectivity opportunities for the GUs.
Fig.~\ref{3D} also demonstrates that the altitude for which the peak value is reached increases as $\lambda_{u}$ decreases: this approach (i) offers better signal quality by increasing the LOS probability (which has, therefore, a more significant impact than the resulting larger path loss), and (ii) produces larger VBs (which generate larger coverage regions).

\subsection{Final Comments and Design Guidelines} 
\label{sub:final_comments_and_design_guidelines}

\begin{figure}[t!]
    \centering
    \includegraphics[width=0.99\columnwidth]{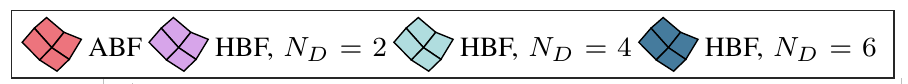}\vspace{-0.3cm}
\end{figure}

\begin{figure}[t!]
    \centering
    \begin{subfigure}[t!]{0.47\textwidth}
    \setlength\fwidth{0.88\textwidth}
    \setlength\fheight{0.8\textwidth}
    \includegraphics[width=0.99\columnwidth]{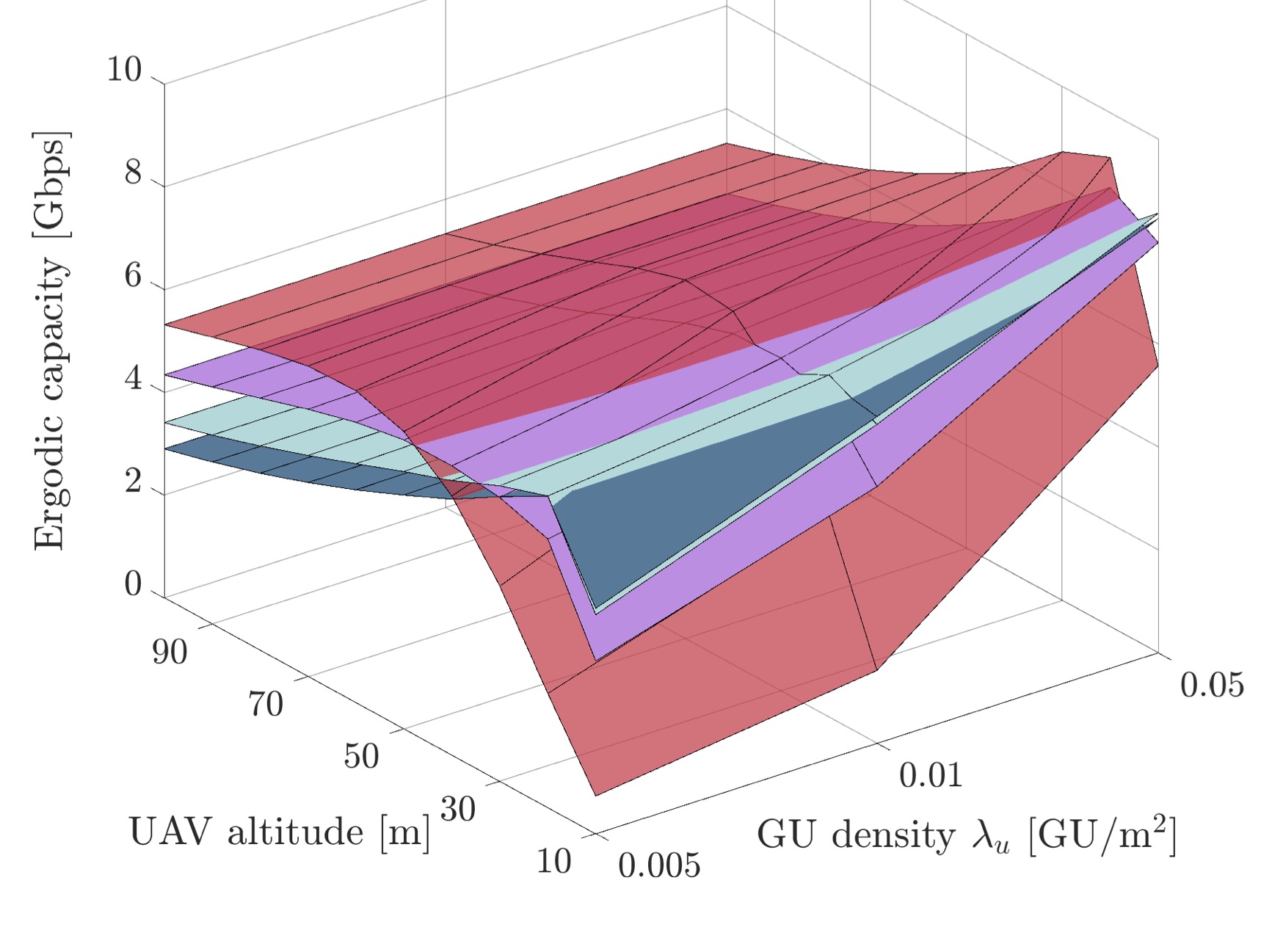}
        \caption{Antenna array size $N_T=9\times9$.}
                \label{3d0.01}
    \end{subfigure}\\ \vspace{0.66cm}
    \begin{subfigure}[t!]{0.47\textwidth}
    \setlength\fwidth{0.88\textwidth}
    \setlength\fheight{0.8\textwidth}
    \includegraphics[width=0.99\columnwidth]{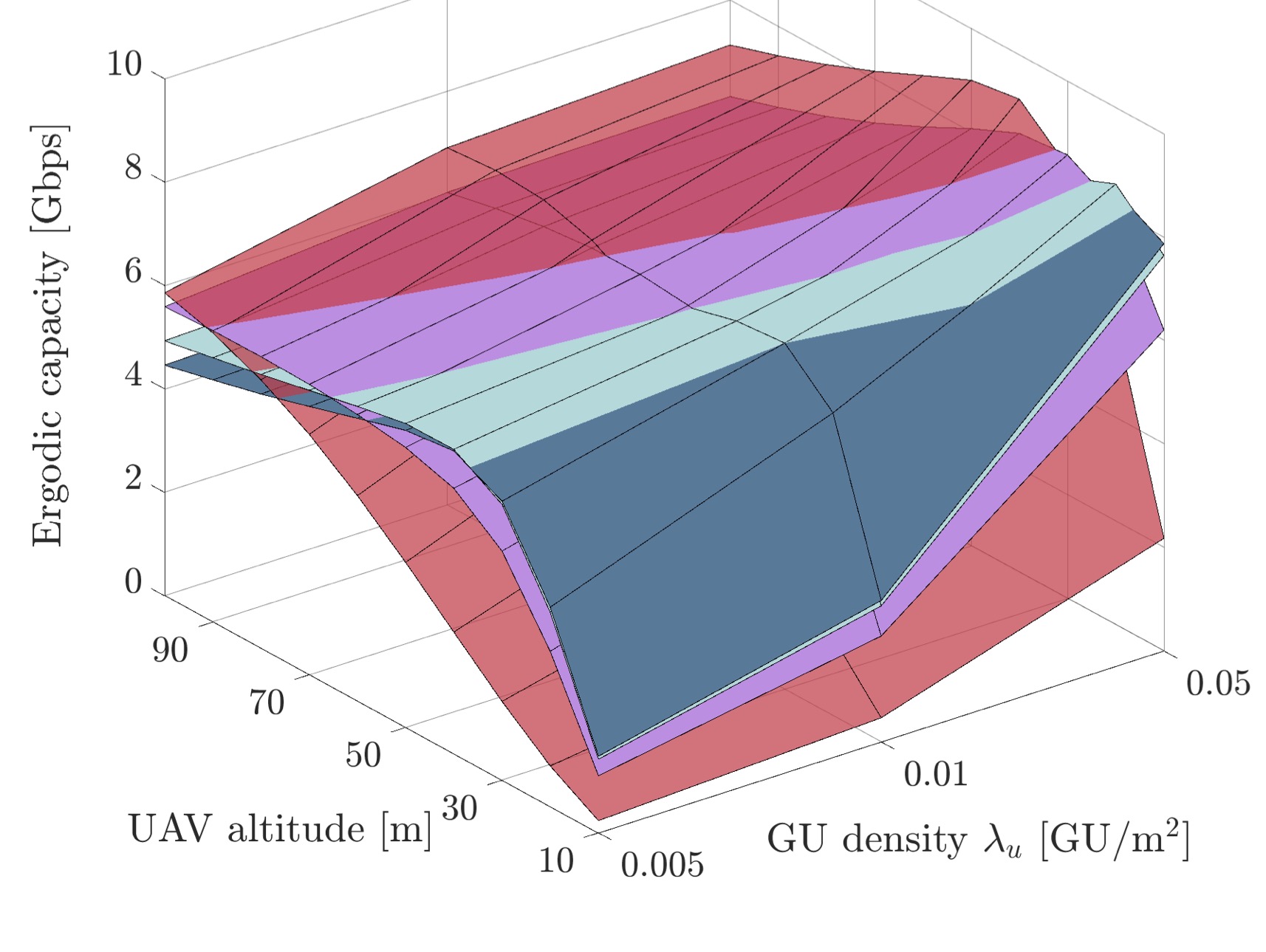}
        \caption{Antenna array size $N_T=17\times17$.}
                \label{3d0.005}
        \end{subfigure}
    \caption{Ergodic capacity of ABF and HBF vs. the deployment altitude, the antenna array size, the number of parallel beams in HBF, and the GU's deployment density.}
    \label{3D}
\end{figure}

Based on the results we presented in the previous subsections, we now provide some guidelines to optimally dimension a UAV-operated network at \gls{mmwave} frequencies  to jointly optimize power consumption and ergodic capacity.
\smallskip

\begin{figure}[t!]
    \centering
    \begin{subfigure}[t!]{0.44\textwidth}
    \centering
    \setlength\fwidth{0.9\columnwidth}
    \setlength\fheight{0.65\columnwidth}
        \definecolor{mycolor1}{RGB}{230, 57, 70}%
\definecolor{mycolor2}{RGB}{205, 143, 229}%
\definecolor{mycolor3}{RGB}{168, 218, 220}%
\definecolor{mycolor4}{RGB}{69, 123, 157}%
\definecolor{mycolor5}{RGB}{29, 53, 87}%

\tikzset{
every pin/.append style={font=\scriptsize, align = left},
}
\begin{tikzpicture}[/pgfplots/tick scale binop=\times]
\pgfplotsset{
tick label style={font=\scriptsize},
label style={font=\scriptsize},
legend  style={font=\scriptsize},
}

\begin{axis}[%
width=0.951\fwidth,
height=\fheight,
at={(0\fwidth,0\fheight)},
scale only axis,
bar shift auto,
xmin=0.509090909090909,
xmax=5.49090909090909,
xtick={1,2,3,4,5},
xticklabels={10,20,30,40,50},
xlabel={UAV altitude [m]},
ymin=0,
ymax=35000000,
ylabel={Power efficiency [bps/W/Hz]},
xlabel style={font=\scriptsize\color{white!15!black}},
ylabel style={font=\scriptsize\color{white!15!black}},
xmajorgrids,
ymajorgrids,
legend style={legend cell align=left, align=left, draw=white!15!black, at={(0.5,1.2)},/tikz/every even column/.append style={column sep=0.15cm},
  anchor=north ,legend columns=-1}
]
\addplot[ybar, bar width=0.18, fill=mycolor1, draw=black, area legend] table[row sep=crcr] {%
1	2245417.81798976\\
2	7271262.79695311\\
3	13605587.5322051\\
4	20088239.2163385\\
5	25846595.1914562\\
};
\addplot[forget plot, color=white!15!black] table[row sep=crcr] {%
0.511111111111111	0\\
5.48888888888889	0\\
};
\addlegendentry{ABF}

\addplot[ybar, bar width=0.18, fill=mycolor2, draw=black, area legend,postaction={pattern=north east lines, opacity=0.5}] table[row sep=crcr] {%
1	10081891.1364147\\
2	23248412.6712955\\
3	29209069.5256695\\
4	30956942.1900526\\
5	31685214.5882546\\
};
\addplot[forget plot, color=white!15!black] table[row sep=crcr] {%
0.511111111111111	0\\
5.48888888888889	0\\
};
\addlegendentry{HBF, $N_D=2$}


\addplot[ybar, bar width=0.18, fill=mycolor3, draw=black, area legend,postaction={pattern=dots, opacity=0.5}] table[row sep=crcr] {%
1	12398564.2008327\\
2	27312691.510851\\
3	32228294.6122204\\
4	31755801.6301029\\
5	30344450.7581794\\
};
\addplot[forget plot, color=white!15!black] table[row sep=crcr] {%
0.511111111111111	0\\
5.48888888888889	0\\
};
\addlegendentry{HBF, $N_D=4$}

\end{axis}
\end{tikzpicture}%
    \caption{$\lambda_{u}=0.005$ GU/m$^2$ and $N_T=13\times13$.}
    \label{fig:ee-low}
    \end{subfigure}\\ \vspace{0.33cm}
    \begin{subfigure}[t!]{0.47\textwidth}
    \centering
    \setlength\fwidth{0.85\columnwidth}
    \setlength\fheight{0.75\columnwidth}
    \includegraphics[width=0.93\textwidth]{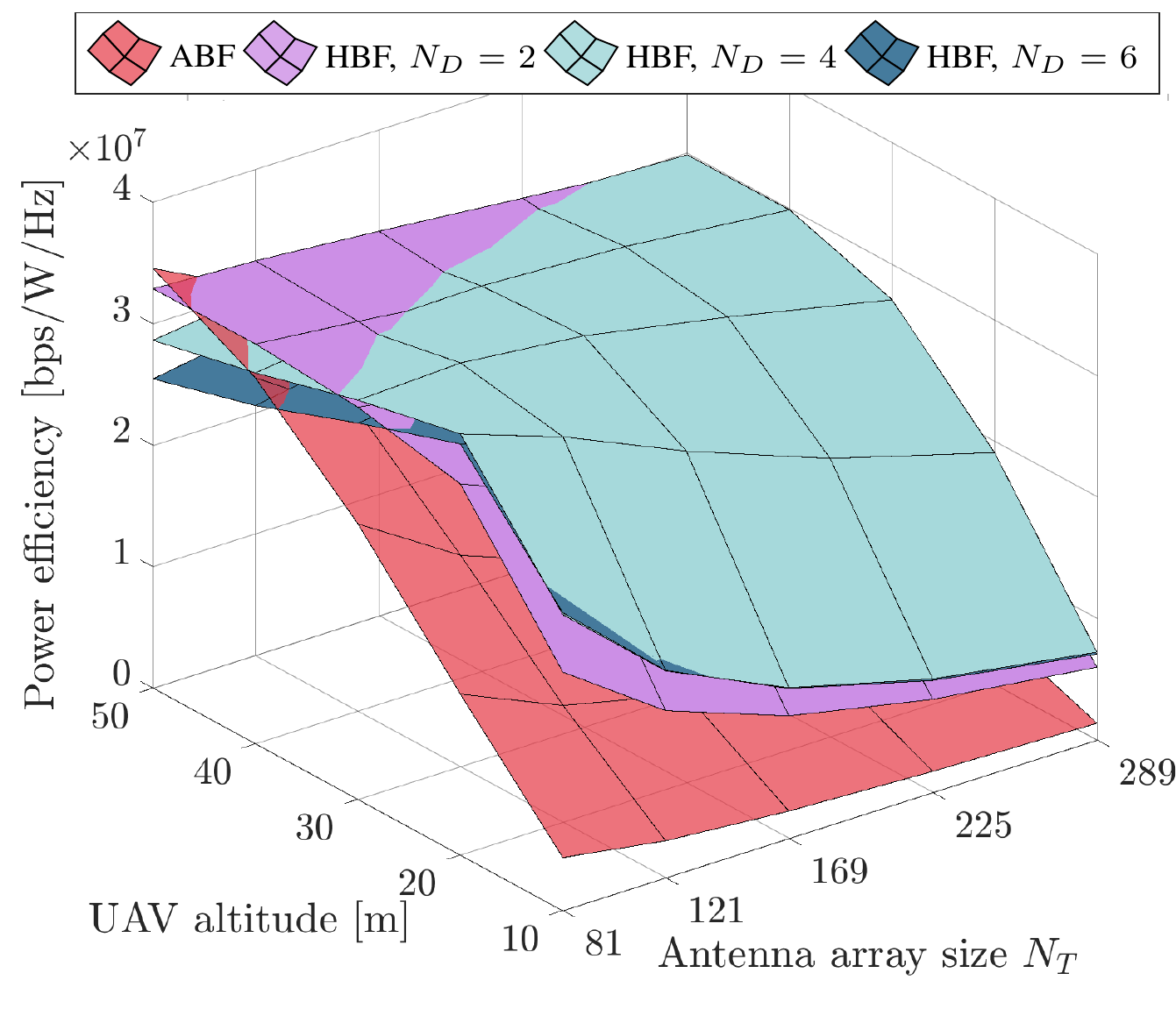}
    \caption{$\lambda_{u}=0.005$ GU/m$^2$.}
    \label{fig:ee-hist}
        \end{subfigure}
    \caption{Power efficiency of ABF and HBF vs. the deployment altitude, the number of parallel beams in HBF, and $N_T$.}
    \label{fig:ee}
\end{figure}
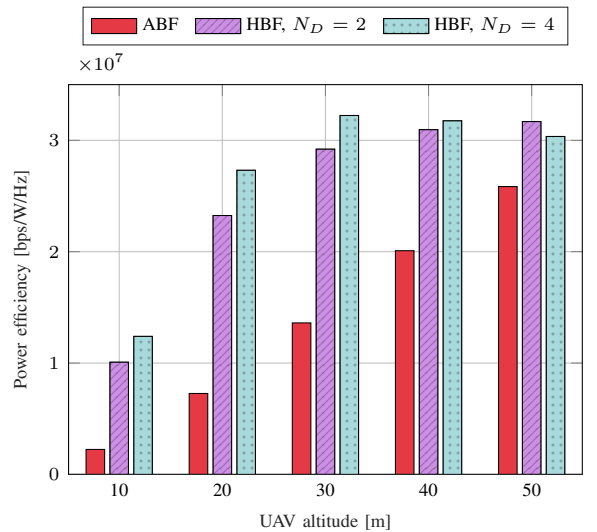
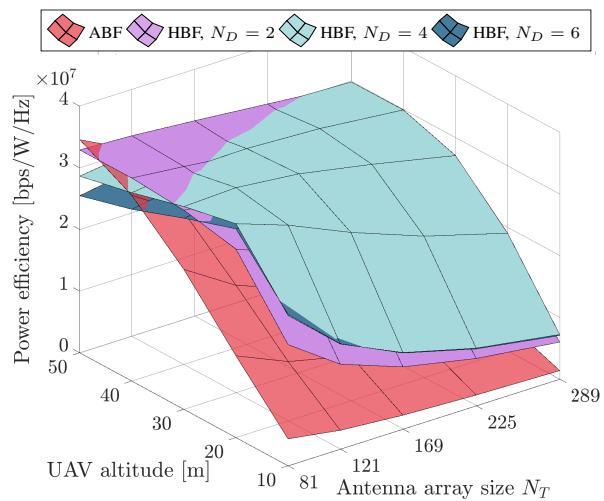

\emph{ABF vs. HBF.} In general, ABF provides the highest ergodic capacity since the  transmit power is concentrated towards a single vertical beam, while HBF can minimize the total power consumption  when $N_D>1$ by reducing the number of UAVs that need to be deployed to achieve full service coverage.
However, it should be noted that the capacity degradation for HBF is quite limited (e.g., only $-6\%$ for $N_T=9\times9$ and $N_D=2$ at $h^*=25$ m, as illustrated in Fig.~\ref{rah_c}), in the face of a performance improvement in terms of power consumption by $60\%$ (as depicted in Fig.~\ref{fig:power-h}), thus making HBF an attractive option.
Moreover, Fig.~\ref{fig:ee} illustrates that HBF guarantees better power efficiency, which is computed as $\tau_{i}/P_C^{i}$, $i\in\{\rm ABF,HBF\}$, in sparse deployments (i.e., $\lambda_u=0.005$ GU/m$^2$, which identify a typical emergency situation in public safety scenarios) for all investigated configurations, by offering the GUs more connectivity opportunities through tilted beams. 
Fig.~\ref{fig:ee-hist} also exemplifies that HBF outperforms ABF when decreasing the UAV altitude, and when increasing the antenna array size (which is required to serve the traffic of the farthest GUs through the resulting higher gain achieved by beamforming). 
\smallskip

\emph{Impact of $h$.}
From a theoretical point of view, it would be preferable to increase the deployment altitude to expand the service region shaped by VBs and TBs on the ground, and consistently reduce the number of UAVs that are required to cover the AoI.
In practice, there exists an optimal value  $h^*$ that maximizes the ergodic capacity:  for $h>h^*$, despite the more likely LOS links, the impact of the increased path loss between the GUs and the serving UAV reduces the overall link budget, thus leading to intermittent connectivity.
Notably, the value of $h^*$ increases as $\lambda_u$ decreases, especially when ABF is implemented. 
For example, Fig.~\ref{3d0.01} shows that, for $N_T=9\times9$, $h^*$  moves from $60$ m when $\lambda_u=0.005$ GU/m$^2$ to $30$ m when $\lambda_u=0.05$ GU/m$^2$.
\smallskip

\emph{Impact of $N_D$ and $N_{RF}$.} HBF permits to form $N_D$ parallel beams to serve multiple GUs~simultaneously from the same UAV station.
This approach (i) guarantees reduced power consumption compared to ABF, and (ii) offers higher ergodic capacity when UAVs hover at low altitude.
Nevertheless, $N_D$ exhibits a maximum, that is inversely proportional to the GU density $\lambda_u$, above which increasing the number of TBs would deteriorate the ergodic capacity (as the transmit power has to be split among more directions simultaneously) with limited performance improvements in terms of power consumption.
Finally, the optimal design would be to set $N_D = N_{RF}$.
\smallskip

\emph{Impact of $N_T$.}
In general, $N_T$ should be reduced as much as possible. 
Not only can this strategy minimize the total power consumption (by reducing the number of electronic components in the RF chain(s)), but it can also improve the ergodic capacity by allowing UAVs to generate larger connectivity regions on the ground. 
However, $N_T$ might still need to be increased, despite the higher power consumption, to provide connectivity (i) when $h$ is large, or (ii) when $\lambda_u$ is small, i.e., when the distance to the serving UAV increases. For example, from Fig.~\ref{rah_c} we can see that the ergodic capacity grows by more than $30\%$ when $N_T$ is increased from $9\times9$ to $17\times17$ at $h=100$ m.

\section{Conclusions and Future Work}
\label{sec:conclusions}
In this work we characterized the ergodic capacity and power consumption of UAV mmWave  networks and investigated the relationship between UAV altitude, GU density, and antenna design.
Our model takes into account the power consumed by both UAV's hovering and communication hardware components, and  derives an analytical expression of the ergodic capacity for ABF and HBF architectures.
We validated our analytical curves via Monte Carlo simulations, and demonstrated that there exists an optimal altitude at which the UAV should be placed to improve network capacity, which depends on the GU density and the adopted beamforming architecture.
Specifically, we showed that HBF always consumes the least power, especially when increasing the antenna array size, with minimum capacity degradation compared to ABF. 
Moreover, the hybrid approach outperforms ABF in terms of capacity too when considering sparse GU deployments and when the UAVs hover at low altitudes (e.g., during take off or when civil flight authorities pose a limit on the height of the UAV).

This work opens up some particularly interesting research directions, such as the definition of a hybrid mechanism able to dynamically identify the recommended network configuration(s) as a function of the UAV swarm size, altitude, and antenna architecture.
We will also investigate how to practically integrate UAVs with other non-terrestrial platforms like \glspl{hap} or satellites to promote ubiquitous and high-capacity global connectivity to on-the-ground devices.

\appendices
\section{Proof of Theorem~\ref{Theorem: rate for VB}}
\label{app:th1}
    With the assumption that the small scale fading is modeled as a Nakagami-m random variable $\gamma_i$ with parameter $\m_{i}$ and $\Omega_{i}$, $i\in\{L,N\}$, for a given constant $x > 0$, the probability $\mathbb{P}[\gamma_{i} > x]$ can be written as
    \begin{align}
    \mathbb{P}[\gamma_{i}> x]&=1-\mathbb{P}[\gamma_{i}\leq x] \notag \\
    &  =1-P\left(m_{i},\frac{m_{i}}{\Omega_{i}}x^{2}\right)    \nonumber\\
    &=1-\frac{u\left(m_{i},\frac{m_{i}}{\Omega_{i}}x^{2}\right)}{\Gamma(m_{i})} \notag \\
    &  =1-\frac{\int_{0}^{\frac{m_{i}}{\Omega_{i}}x}l^{m_{i}-1}e^{-l}dl}{(m_{i}-1)!}
    \label{eq:th1-naka}
    \end{align}
    where $P(m_{i},\frac{m_{i}}{\Omega_{i}}x^{2})$ is the (regularized) incomplete gamma function, and $u(m_{i},\frac{m_{i}}{\Omega_{i}}x^{2})$ is the lower incomplete gamma function.
    Let $r_1$ be the 2D distance between the UAV and the closest GU $\in \Phi_u$ within the VB, whose \gls{pdf} is expressed in Eq.~\eqref{eq:f_r_1}. The ergodic capacity experienced from the GU in the VB can be rewritten from Eq.~\eqref{eq:SNR} as
    \begin{align}
    &\tau^{\rm VB}  \overset{\Delta}{=}\mathbb{E}_{\Phi_u}\left[B \log_2(1+\text{SNR}(r)) \right]  \nonumber\\[10pt]
    &=\int_{0}^{r_{\rm VB}} B  f_{r1}^{\rm VB}(r)  \mathbb{E}_{\Phi_u}\left\{\log_2 \left(1+\frac{P_{t}G\gamma_{L}p_{L}(r)\ell_{L}(r)}{N_{D} \cdot\mathrm{NF} \cdot\sigma ^{2}}\right.\right. \notag \\[10pt]
    & \left. \left. +\frac{P_tG\gamma_{N}p_{N}(r)\ell_{N}(r)}{N_{D}\cdot \mathrm{NF} \cdot \sigma ^{2}}\right) \right\}  \mathrm{d}r  \nonumber\\[10pt] 
    &\overset{(a)}{=}\int_{0}^{r_{\rm VB}} \int_{0}^{\infty} B  f_{r_1}^{\rm VB}(r)  \mathbb{P}[\gamma _{L}>\beta(t,\gamma_{N})]  \mathrm{d}t \mathrm{d}r    \nonumber\\[10pt]
    &\overset{(b)}{=}\int_{0}^{r_{\rm VB}} \int_{0}^{\infty}  B  f_{r_1}^{\rm VB}(r)  \mathbb{E}_{\Phi _{u,N}} \left[ 1- \frac{\int_{0}^{\Delta_L}l^{m_{L}-1}e^{-l}dl}{(m_{L}-1)!} \right]  \mathrm{d}t  \mathrm{d}r \notag
    \end{align}
        \medmuskip=-1mu
\thickmuskip=-1mu
    \begin{align}
    &\overset{(c)}{=}\int_{0}^{r_{\rm VB}}\hspace{-0.15cm} \int_{0}^{\infty}  \left(\hspace{-0.1cm}1+   \frac{\mathbb{E}_{\Phi _{u,N}}\hspace{-0.15cm}\left[\left(\frac{l^{m_L}}{l}+\sum\limits_{j=2}^{m_{L}}\frac{\Gamma(m_L)}{(m_L-j)!} l^{m_{L}-j}\right)\hspace{-0.1cm}e^{-l} \Big|_{0}^{\Delta_L}\right]  }{(m_{L}-1)!} \hspace{-0.05cm}\right) \notag \\
    &\cdot B  f_{r_1}^{\rm VB}(r) \mathrm{d}t \mathrm{d}r  \notag
        \end{align}
        \medmuskip=4mu
\thickmuskip=4mu
\vspace{-0.66cm}
    \begin{align}
    &=\int_{0}^{r_{\rm VB}} \int_{0}^{\infty}  \frac{\mathbb{E}_{\Phi _{u,N}} \left[\left(l^{m_L-1}+\sum\limits_{j=2}^{m_{L}}\frac{\Gamma(m_L)}{(m_L-j)!} l^{m_{L}-j}\right) e^{-\Delta_L} \right]  }{(m_{L}-1)!}\notag \\[10pt]
    &\cdot B  f_{r_1}^{\rm VB}(r) \mathrm{d}t \mathrm{d}r  \nonumber\\    
    &=B\int_{0}^{r_{\rm VB}} \int_{0}^{\infty} \int_{0}^{\infty} \frac{\left(l^{m_L-1}+\sum\limits_{j=2}^{m_{L}}\frac{\Gamma(m_L)}{(m_L-j)!} l^{m_{L}-j}\right) e^{-\Delta_L}   }{(m_{L}-1)!} \notag \\
    &\cdot  f_{r_1}^{\rm VB}(r)  p_\gamma(\gamma_N) \mathrm{d}\gamma_{N} \mathrm{d}t \mathrm{d}r,
    \label{eq:proof}
    \end{align}

    where $(a)$ follows the fact that, for a positive random variable X, $\mathbb{E}(X)=\int_{t>0}\mathbb{P}(X>t)\mathrm{d} t$ \cite{6042301}, $(b)$ is obtained from Eq.~\eqref{eq:th1-naka}, and $(c)$ derives from 
    \begin{align}
&\int_{0}^{\Delta_L}l^{m_{L}-1}e^{-l}dl \notag \\
&= -(l^{m_{L}-1}+\sum_{j=2}^{m_{L}} \frac{\Gamma(m_{L})}{(m_{L}-j)} l^{m_{L}-j}) e^{-l} \Big|_{0}^{\Delta_L}.
\end{align}
   In Eq.~\eqref{eq:proof}, $p_\gamma(\gamma_N)$ is the probability distribution of the Nakagami-m small-scale fading for NLOS propagation, and we assumed
   \medmuskip=2mu
\thickmuskip=2mu
\begin{align}
\beta(t,\gamma_{N})= \frac{(e^{t\ln2}-1)N_{D}\mathrm{NF}\sigma ^{2}-P_{t}G\gamma _{N}p_{N}(r)\ell_{N}(r)}{P_{t}Gp_{L}(r)\ell_{L}(r)},
\end{align}
   \medmuskip=6mu
\thickmuskip=6mu
\begin{equation}
\Delta_L =  \frac{m_{L}}{\Omega_{L}}\beta(t,\gamma_{N}).
\end{equation}

%
%
%
%

\bibliographystyle{IEEEtran}
\bibliography{mybib}

\begin{thebibliography}{10}
\providecommand{\url}[1]{#1}
\csname url@samestyle\endcsname
\providecommand{\newblock}{\relax}
\providecommand{\bibinfo}[2]{#2}
\providecommand{\BIBentrySTDinterwordspacing}{\spaceskip=0pt\relax}
\providecommand{\BIBentryALTinterwordstretchfactor}{4}
\providecommand{\BIBentryALTinterwordspacing}{\spaceskip=\fontdimen2\font plus
\BIBentryALTinterwordstretchfactor\fontdimen3\font minus
  \fontdimen4\font\relax}
\providecommand{\BIBforeignlanguage}[2]{{%
\expandafter\ifx\csname l@#1\endcsname\relax
\typeout{** WARNING: IEEEtran.bst: No hyphenation pattern has been}%
\typeout{** loaded for the language `#1'. Using the pattern for}%
\typeout{** the default language instead.}%
\else
\language=\csname l@#1\endcsname
\fi
#2}}
\providecommand{\BIBdecl}{\relax}
\BIBdecl

\bibitem{khan2016multi}
F.~Khan, ``Multi-comm-core architecture for terabit-per-second wireless,''
  \emph{IEEE Communications Magazine}, vol.~54, no.~4, pp. 124--129, Apr 2016.

\bibitem{giordani2020towards}
M.~{Giordani}, M.~{Polese}, M.~{Mezzavilla}, S.~{Rangan}, and M.~{Zorzi},
  ``{Toward 6G Networks: Use Cases and Technologies},'' \emph{IEEE
  Communications Magazine}, vol.~58, no.~3, Mar. 2020.

\bibitem{chaoub20206g}
\BIBentryALTinterwordspacing
A.~Chaoub, M.~Giordani, B.~Lall, V.~Bhatia, A.~Kliks, L.~Mendes, K.~Rabie,
  H.~Saarnisaari, A.~Singhal, N.~Zhang \emph{et~al.}, ``{6G for Bridging the
  Digital Divide: Wireless Connectivity to Remote Areas},'' 2020,
  \textit{submitted to the IEEE Wireless Communications Magazine}, 2020.
  [Online]. Available: \url{https://arxiv.org/abs/2009.04175}
\BIBentrySTDinterwordspacing

\bibitem{giordani2019non}
\BIBentryALTinterwordspacing
M.~Giordani and M.~Zorzi, ``{Non-Terrestrial Networks in the 6G Era: Challenges
  and Opportunities},'' \textit{submitted to IEEE Network}, 2020. [Online].
  Available: \url{https://arxiv.org/abs/1912.10226}
\BIBentrySTDinterwordspacing

\bibitem{giordani2020satellite}
------, ``{Satellite Communication at Millimeter Waves: a Key Enabler of the 6G
  Era},'' \emph{IEEE International Conference on Computing, Networking and
  Communications (ICNC)}, Feb 2020.

\bibitem{li2018uav}
B.~Li, Z.~Fei, and Y.~Zhang, ``{UAV communications for 5G and beyond: Recent
  advances and future trends},'' \emph{IEEE Internet of Things Journal},
  vol.~6, no.~2, pp. 2241--2263, Apr 2018.

\bibitem{7470933}
Y.~{Zeng}, R.~{Zhang}, and T.~J. {Lim}, ``{Wireless communications with
  unmanned aerial vehicles: opportunities and challenges},'' \emph{IEEE
  Communications Magazine}, vol.~54, no.~5, pp. 36--42, May 2016.

\bibitem{8764406}
L.~{Zhang}, H.~{Zhao}, S.~{Hou}, Z.~{Zhao}, H.~{Xu}, X.~{Wu}, Q.~{Wu}, and
  R.~{Zhang}, ``{A Survey on 5G Millimeter Wave Communications for UAV-Assisted
  Wireless Networks},'' \emph{IEEE Access}, vol.~7, pp. 117\,460--117\,504, Jul
  2019.

\bibitem{8758340}
X.~{Li}, H.~{Yao}, J.~{Wang}, X.~{Xu}, C.~{Jiang}, and L.~{Hanzo}, ``{A
  Near-Optimal UAV-Aided Radio Coverage Strategy for Dense Urban Areas},''
  \emph{IEEE Transactions on Vehicular Technology}, vol.~68, no.~9, pp.
  9098--9109, Jul 2019.

\bibitem{7959169}
M.~{Xiao}, S.~{Mumtaz}, Y.~{Huang}, L.~{Dai}, Y.~{Li}, M.~{Matthaiou}, G.~K.
  {Karagiannidis}, E.~{Björnson}, K.~{Yang}, C.~{I}, and A.~{Ghosh},
  ``{Millimeter Wave Communications for Future Mobile Networks},'' \emph{IEEE
  Journal on Selected Areas in Communications}, vol.~35, no.~9, pp. 1909--1935,
  Jun 2017.

\bibitem{8482308}
M.~{Gapeyenko}, V.~{Petrov}, D.~{Moltchanov}, S.~{Andreev}, N.~{Himayat}, and
  Y.~{Koucheryavy}, ``{Flexible and Reliable UAV-Assisted Backhaul Operation in
  5G mmWave Cellular Networks},'' \emph{IEEE Journal on Selected Areas in
  Communications}, vol.~36, no.~11, pp. 2486--2496, Oct 2018.

\bibitem{8999435}
N.~{Tafintsev}, D.~{Moltchanov}, M.~{Gerasimenko}, M.~{Gapeyenko}, J.~{Zhu},
  S.~{Yeh}, N.~{Himayat}, S.~{Andreev}, Y.~{Koucheryavy}, and M.~{Valkama},
  ``{Aerial Access and Backhaul in mmWave B5G Systems: Performance Dynamics and
  Optimization},'' \emph{IEEE Communications Magazine}, vol.~58, no.~2, pp.
  93--99, Feb 2020.

\bibitem{giordani2019tutorial}
M.~{Giordani}, M.~{Polese}, A.~{Roy}, D.~{Castor}, and M.~{Zorzi}, ``{A
  Tutorial on Beam Management for 3GPP NR at mmWave Frequencies},'' \emph{IEEE
  Communications Surveys Tutorials}, vol.~21, no.~1, pp. 173--196, Firstquarter
  2019.

\bibitem{sun2014mimo}
S.~Sun, T.~S. Rappaport, R.~W. Heath, A.~Nix, and S.~Rangan, ``{MIMO for
  millimeter-wave wireless communications: beamforming, spatial multiplexing,
  or both?}'' \emph{IEEE Communications Magazine}, vol.~52, no.~12, pp.
  110--121, Dec. 2014.

\bibitem{7961162}
C.~G. {Tsinos}, S.~{Maleki}, S.~{Chatzinotas}, and B.~{Ottersten}, ``On the
  energy-efficiency of hybrid analog–digital transceivers for single- and
  multi-carrier large antenna array systems,'' \emph{IEEE Journal on Selected
  Areas in Communications}, vol.~35, no.~9, pp. 1980--1995, Jun. 2017.

\bibitem{xiao2020unmanned}
Z.~{Xiao}, H.~{Dong}, L.~{Bai}, D.~O. {Wu}, and X.~{Xia}, ``{Unmanned Aerial
  Vehicle Base Station (UAV-BS) Deployment With Millimeter-Wave Beamforming},''
  \emph{IEEE Internet of Things Journal}, vol.~7, no.~2, pp. 1336--1349, Feb
  2020.

\bibitem{sohrabi2017hybrid}
F.~Sohrabi and W.~Yu, ``{Hybrid analog and digital beamforming for mmWave OFDM
  large-scale antenna arrays},'' \emph{IEEE Journal on Selected Areas in
  Communications}, vol.~35, no.~7, pp. 1432--1443, Apr 2017.

\bibitem{molisch2017hybrid}
A.~F. Molisch, V.~V. Ratnam, S.~Han, Z.~Li, S.~L.~H. Nguyen, L.~Li, and
  K.~Haneda, ``{Hybrid beamforming for massive MIMO: A survey},'' \emph{IEEE
  Communications Magazine}, vol.~55, no.~9, pp. 134--141, Sep 2017.

\bibitem{ravi2016downlink}
V.~V.~C. Ravi and H.~S. Dhillon, ``{Downlink coverage probability in a finite
  network of unmanned aerial vehicle (UAV) base stations},'' in \emph{IEEE 17th
  International Workshop on Signal Processing Advances in Wireless
  Communications (SPAWC)}, 2016.

\bibitem{chetlur2017downlink}
V.~V. Chetlur and H.~S. Dhillon, ``{Downlink coverage analysis for a finite 3-D
  wireless network of unmanned aerial vehicles},'' \emph{IEEE Transactions on
  Communications}, vol.~65, no.~10, pp. 4543--4558, Oct. 2017.

\bibitem{yi2019modeling}
W.~Yi, Y.~Liu, M.~Elkashlan, and A.~Nallanathan, ``{Modeling and coverage
  analysis of downlink UAV networks with mmWave communications},'' in
  \emph{2019 IEEE International Conference on Communications Workshops (ICC
  Workshops)}.\hskip 1em plus 0.5em minus 0.4em\relax IEEE, 2019, pp. 1--6.

\bibitem{boschiero2020coverage}
M.~{Boschiero}, M.~{Giordani}, M.~{Polese}, and M.~{Zorzi}, ``{Coverage
  Analysis of UAVs in Millimeter Wave Networks: A Stochastic Geometry
  Approach},'' in \emph{2020 International Wireless Communications and Mobile
  Computing (IWCMC)}, 2020, pp. 351--357.

\bibitem{di2015energy}
C.~Di~Franco and G.~Buttazzo, ``{Energy-aware coverage path planning of
  UAVs},'' in \emph{IEEE International Conference on Autonomous Robot Systems
  and Competitions}, 2015.

\bibitem{naqvi2018energy}
S.~Naqvi, J.~Chakareski, N.~Mastronarde, J.~Xu, F.~Afghah, and A.~Razi,
  ``{Energy efficiency analysis of UAV-assisted mmWave HetNets},'' in
  \emph{IEEE International Conference on Communications (ICC)}, 2018.

\bibitem{xiao2016enabling}
Z.~Xiao, P.~Xia, and X.~Xia, ``Enabling {UAV} cellular with millimeter-wave
  communication: Potentials and approaches,'' \emph{IEEE Communications
  Magazine}, vol.~54, no.~5, pp. 66--73, May 2016.

\bibitem{xia2019millimeter}
W.~Xia, M.~Polese, M.~Mezzavilla, G.~Loianno, S.~Rangan, and M.~Zorzi,
  ``{Millimeter Wave Remote UAV Control and Communications for Public Safety
  Scenarios},'' in \emph{Proc. of the 1st Intl. Workshop on Internet of
  Autonomous Unmanned Vehicles}, ser. IAUV '19, Boston, MA, 2019.

\bibitem{zhao2018channel}
J.~Zhao, G.~Gao, L.~Kuang, Q.~Wu, and W.~Jia, ``Channel tracking with flight
  control system for {UAV} {mmWave} {MIMO} communications,'' \emph{IEEE
  Communications Letters}, vol.~22, no.~6, pp. 1224--1227, Jun. 2018.

\bibitem{azari2017coverage}
M.~M. Azari, Y.~Murillo, O.~Amin, F.~Rosas, M.-S. Alouini, and S.~Pollin,
  ``{Coverage maximization for a poisson field of drone cells},'' in \emph{IEEE
  28th Annual International Symposium on Personal, Indoor, and Mobile Radio
  Communications (PIMRC)}, 2017.

\bibitem{liu2018performance}
C.~Liu, M.~Ding, C.~Ma, Q.~Li, Z.~Lin, and Y.-C. Liang, ``{Performance analysis
  for practical unmanned aerial vehicle networks with LoS/NLoS
  transmissions},'' in \emph{IEEE International Conference on Communications
  Workshops (ICC Workshops)}, 2018.

\bibitem{8856258}
W.~{Yi}, Y.~{Liu}, E.~{Bodanese}, A.~{Nallanathan}, and G.~K. {Karagiannidis},
  ``{A Unified Spatial Framework for UAV-Aided MmWave Networks},'' \emph{IEEE
  Transactions on Communications}, vol.~67, no.~12, pp. 8801--8817, Oct 2019.

\bibitem{8876702}
X.~{Wang} and M.~C. {Gursoy}, ``{Coverage Analysis for Energy-Harvesting
  UAV-Assisted mmWave Cellular Networks},'' \emph{IEEE Journal on Selected
  Areas in Communications}, vol.~37, no.~12, pp. 2832--2850, Oct 2019.

\bibitem{colpaert2018aerial}
A.~Colpaert, E.~Vinogradov, and S.~Pollin, ``{Aerial coverage analysis of
  cellular systems at LTE and mmwave frequencies using 3D city models},''
  \emph{Sensors}, vol.~18, no.~12, p. 4311, Dec 2018.

\bibitem{9115248}
L.~{Zhu}, J.~{Zhang}, Z.~{Xiao}, X.~{Cao}, X.~{Xia}, and R.~{Schober},
  ``{Millimeter-Wave Full-Duplex UAV Relay: Joint Positioning, Beamforming, and
  Power Control},'' \emph{IEEE Journal on Selected Areas in Communications},
  vol.~38, no.~9, pp. 2057--2073, Jan 2020.

\bibitem{8907440}
Z.~{Xiao}, H.~{Dong}, L.~{Bai}, D.~O. {Wu}, and X.~{Xia}, ``{Unmanned Aerial
  Vehicle Base Station (UAV-BS) Deployment With Millimeter-Wave Beamforming},''
  \emph{IEEE Internet of Things Journal}, vol.~7, no.~2, pp. 1336--1349, Feb
  2020.

\bibitem{mozaffari2015drone}
M.~Mozaffari, W.~Saad, M.~Bennis, and M.~Debbah, ``{Drone small cells in the
  clouds: Design, deployment and performance analysis},'' in \emph{IEEE Global
  Communications Conference (GLOBECOM)}, 2015.

\bibitem{yang2018energy}
D.~Yang, Q.~Wu, Y.~Zeng, and R.~Zhang, ``{Energy tradeoff in ground-to-UAV
  communication via trajectory design},'' \emph{IEEE Transactions on Vehicular
  Technology}, vol.~67, no.~7, pp. 6721--6726, Mar. 2018.

\bibitem{zeng2019energy}
Y.~Zeng, J.~Xu, and R.~Zhang, ``{Energy minimization for wireless communication
  with rotary-wing UAV},'' \emph{IEEE Transactions on Wireless Communications},
  vol.~18, no.~4, pp. 2329--2345, Apr. 2019.

\bibitem{orhan2015low}
O.~Orhan, E.~Erkip, and S.~Rangan, ``{Low power analog-to-digital conversion in
  millimeter wave systems: Impact of resolution and bandwidth on
  performance},'' in \emph{Information Theory and Applications Workshop (ITA)},
  2015.

\bibitem{buzzi2018energy}
S.~Buzzi and C.~D'Andrea, ``{Energy efficiency and asymptotic performance
  evaluation of beamforming structures in doubly massive MIMO mmWave
  systems},'' \emph{IEEE Transactions on Green Communications and Networking},
  vol.~2, no.~2, pp. 385--396, Jan. 2018.

\bibitem{miao2019position}
W.~{Miao}, C.~{Luo}, G.~{Min}, L.~{Wu}, T.~{Zhao}, and Y.~{Mi},
  ``{Position-Based Beamforming Design for UAV Communications in LTE
  Networks},'' in \emph{IEEE International Conference on Communications (ICC)},
  2019.

\bibitem{yang2019beam}
L.~Yang and W.~Zhang, ``{Beam tracking and optimization for UAV
  communications},'' \emph{IEEE Transactions on Wireless Communications},
  vol.~18, no.~11, pp. 5367--5379, Nov. 2019.

\bibitem{miao2019lightweight}
W.~{Miao}, C.~{Luo}, G.~{Min}, and Z.~{Zhao}, ``Lightweight 3-d beamforming
  design in 5g uav broadcasting communications,'' \emph{IEEE Transactions on
  Broadcasting}, vol.~66, no.~2, pp. 515--524, Jun 2020.

\bibitem{li2020reconfigurable}
S.~Li, B.~Duo, X.~Yuan, Y.-C. Liang, and M.~Di~Renzo, ``{Reconfigurable
  intelligent surface assisted UAV communication: Joint trajectory design and
  passive beamforming},'' \emph{IEEE Wireless Communications Letters}, vol.~9,
  no.~5, pp. 716--720, May 2020.

\bibitem{yu2019low}
Q.~Yu, C.~Han, J.~Wang, and L.~Bai, ``{Low Complexity Hybrid Beamforming for
  MmWave-UAV Communication Systems with a Pre-defined Codebook},'' in
  \emph{Proceedings of the 2nd International Conference on Control and Computer
  Vision}, 2019.

\bibitem{ren2019machine}
H.~Ren, L.~Li, W.~Xu, W.~Chen, and Z.~Han, ``{Machine learning-based hybrid
  precoding with robust error for UAV mmWave massive MIMO},'' in \emph{IEEE
  International Conference on Communications (ICC)}, 2019.

\bibitem{al2014}
{ITU}, ``{Propagation data and prediction methods for the design of terrestrial
  broadband millimetric radio access systems},'' P.1410-2, 2003.

\bibitem{al2014optimal}
A.~Al-Hourani, S.~Kandeepan, and S.~Lardner, ``{Optimal LAP altitude for
  maximum coverage},'' \emph{IEEE Wireless Communications Letters}, vol.~3,
  no.~6, pp. 569--572, Jul. 2014.

\bibitem{mozaffari2016unmanned}
M.~Mozaffari, W.~Saad, M.~Bennis, and M.~Debbah, ``{Unmanned aerial vehicle
  with underlaid device-to-device communications: Performance and tradeoffs},''
  \emph{IEEE Transactions on Wireless Communications}, vol.~15, no.~6, pp.
  3949--3963, Feb. 2016.

\bibitem{6515173}
T.~S. {Rappaport}, S.~{Sun}, R.~{Mayzus}, H.~{Zhao}, Y.~{Azar}, K.~{Wang},
  G.~N. {Wong}, J.~K. {Schulz}, M.~{Samimi}, and F.~{Gutierrez}, ``{Millimeter
  Wave Mobile Communications for 5G Cellular: It Will Work!}'' \emph{IEEE
  Access}, vol.~1, pp. 335--349, May 2013.

\bibitem{zhu2018secrecy}
Y.~{Zhu}, G.~{Zheng}, and M.~{Fitch}, ``{Secrecy Rate Analysis of UAV-Enabled
  mmWave Networks Using Matern Hardcore Point Processes},'' \emph{IEEE Journal
  on Selected Areas in Communications}, vol.~36, no.~7, pp. 1397--1409, Jul.
  2018.

\bibitem{nakagami1960m}
M.~Nakagami, ``The m-distribution—a general formula of intensity distribution
  of rapid fading,'' in \emph{Statistical methods in radio wave
  propagation}.\hskip 1em plus 0.5em minus 0.4em\relax Elsevier, 1960, pp.
  3--36.

\bibitem{dabiri2020analytical}
M.~T. Dabiri, H.~Safi, S.~Parsaeefard, and W.~Saad, ``{Analytical channel
  models for millimeter wave UAV networks under hovering fluctuations},''
  \emph{IEEE Transactions on Wireless Communications}, vol.~19, no.~4, pp.
  2868--2883, Feb. 2020.

\bibitem{yu2017coverage}
X.~Yu, J.~Zhang, M.~Haenggi, and K.~B. Letaief, ``{Coverage analysis for
  millimeter wave networks: The impact of directional antenna arrays},''
  \emph{IEEE Journal on Selected Areas in Communications}, vol.~35, no.~7, pp.
  1498--1512, Jul 2017.

\bibitem{balanis2016antenna}
C.~A. Balanis, \emph{{Antenna theory: analysis and design}}.\hskip 1em plus
  0.5em minus 0.4em\relax John Wiley \& Sons, 2016.

\bibitem{7991310}
Z.~{Liu}, R.~{Sengupta}, and A.~{Kurzhanskiy}, ``{A power consumption model for
  multi-rotor small unmanned aircraft systems},'' in \emph{International
  Conference on Unmanned Aircraft Systems (ICUAS)}, 2017.

\bibitem{8057288}
W.~B. {Abbas}, F.~{Gomez-Cuba}, and M.~{Zorzi}, ``{Millimeter Wave Receiver
  Efficiency: A Comprehensive Comparison of Beamforming Schemes With Low
  Resolution ADCs},'' \emph{IEEE Transactions on Wireless Communications},
  vol.~16, no.~12, pp. 8131--8146, Dec 2017.

\bibitem{8333733}
L.~N. {Ribeiro}, S.~{Schwarz}, M.~{Rupp}, and A.~L.~F. {de Almeida}, ``{Energy
  Efficiency of mmWave Massive MIMO Precoding With Low-Resolution DACs},''
  \emph{IEEE Journal of Selected Topics in Signal Processing}, vol.~12, no.~2,
  pp. 298--312, May 2018.

\bibitem{6042301}
J.~G. {Andrews}, F.~{Baccelli}, and R.~K. {Ganti}, ``{A Tractable Approach to
  Coverage and Rate in Cellular Networks},'' \emph{IEEE Transactions on
  Communications}, vol.~59, no.~11, pp. 3122--3134, Nov. 2011.

\bibitem{7313829}
D.~{Pepe} and D.~{Zito}, ``{A 78.8–92.8 GHz 4-bit 0–360 active phase
  shifter in 28nm FDSOI CMOS with 2.3 dB average peak gain},'' in \emph{41st
  European Solid-State Circuits Conference (ESSCIRC)}, 2015.

\bibitem{4684642}
K.~{Scheir}, S.~{Bronckers}, J.~{Borremans}, P.~{Wambacq}, and Y.~{Rolain},
  ``{A 52 GHz Phased-Array Receiver Front-End in 90 nm Digital CMOS},''
  \emph{IEEE Journal of Solid-State Circuits}, vol.~43, no.~12, pp. 2651--2659,
  Dec. 2008.

\bibitem{yu201060}
Y.~Yu, P.~G. Baltus, A.~de~Graauw, E.~van~der Heijden, C.~S. Vaucher, and A.~H.
  van Roermund, ``{A 60 GHz phase shifter integrated with LNA and PA in 65 nm
  CMOS for phased array systems},'' \emph{IEEE Journal of Solid-State
  Circuits}, vol.~45, no.~9, pp. 1697--1709, Aug 2010.

\bibitem{kraemer2011design}
M.~Kraemer, D.~Dragomirescu, and R.~Plana, ``{Design of a very low-power,
  low-cost 60 GHz receiver front-end implemented in 65 nm CMOS technology},''
  \emph{International Journal of Microwave and Wireless Technologies}, vol.~3,
  no.~2, pp. 131--138, Apr. 2011.

\bibitem{7913628}
X.~{Yu}, J.~{Zhang}, M.~{Haenggi}, and K.~B. {Letaief}, ``{Coverage Analysis
  for Millimeter Wave Networks: The Impact of Directional Antenna Arrays},''
  \emph{IEEE Journal on Selected Areas in Communications}, vol.~35, no.~7, pp.
  1498--1512, Apr 2017.

\bibitem{Marcu:EECS-2011-132}
C.~Marcu, ``{LO Generation and Distribution for 60GHz Phased Array
  Transceivers},'' Ph.D. dissertation, EECS Department, University of
  California, Berkeley, Dec 2011.

\bibitem{rangan2013energy}
S.~Rangan, T.~Rappaport, E.~Erkip, Z.~Latinovic, M.~R. Akdeniz, and Y.~Liu,
  ``{Energy efficient methods for millimeter wave picocellular systems},'' in
  \emph{IEEE Communication Theory Workshop}, 2013.

\bibitem{mendez2016hybrid}
R.~M{\'e}ndez-Rial, C.~Rusu, N.~Gonz{\'a}lez-Prelcic, A.~Alkhateeb, and R.~W.
  Heath, ``{Hybrid MIMO architectures for millimeter wave communications: Phase
  shifters or switches?}'' \emph{IEEE Access}, vol.~4, pp. 247--267, Jan. 2016.

\bibitem{8335329}
Y.~{Zhu}, G.~{Zheng}, and M.~{Fitch}, ``Secrecy rate analysis of uav-enabled
  mmwave networks using matérn hardcore point processes,'' \emph{IEEE Journal
  on Selected Areas in Communications}, vol.~36, no.~7, pp. 1397--1409, 2018.

\bibitem{6834753}
M.~R. {Akdeniz}, Y.~{Liu}, M.~K. {Samimi}, S.~{Sun}, S.~{Rangan}, T.~S.
  {Rappaport}, and E.~{Erkip}, ``{Millimeter Wave Channel Modeling and Cellular
  Capacity Evaluation},'' \emph{IEEE Journal on Selected Areas in
  Communications}, vol.~32, no.~6, pp. 1164--1179, Jun. 2014.

\bibitem{navolio2011report}
M.~Navolio, ``{Public Safety Wireless Data Network Requirements Project},'' in
  \emph{Needs Assessment Report}, 2011.

\bibitem{giordani2017improved}
M.~{Giordani} and M.~{Zorzi}, ``{Improved user tracking in 5G millimeter wave
  mobile networks via refinement operations},'' in \emph{16th Annual
  Mediterranean Ad Hoc Networking Workshop (Med-Hoc-Net)}, 2017.

\bibitem{tracy20155g}
\BIBentryALTinterwordspacing
T.~McElvaney, ``{5G: From a Public Safety Perspective},'' 2015. [Online].
  Available: \url{http:
  //www.atis.org/5g/presentations/5G_PublicSafety_TMcElvaney.pdf}
\BIBentrySTDinterwordspacing

\end{thebibliography}

%
%
%
%

\end{document}